\documentclass[%
 aip,
 amsmath,amssymb,
 reprint,%
]{revtex4-1}

\usepackage{graphicx}
\usepackage{dcolumn}
\usepackage{bm}

\usepackage[utf8]{inputenc}
\usepackage[T1]{fontenc}
\usepackage{mathptmx}
\usepackage{etoolbox}

\usepackage{verbatim}

\usepackage[english]{babel} 

\makeatletter
\def\@email#1#2{%
 \endgroup
 \patchcmd{\titleblock@produce}
  {\frontmatter@RRAPformat}
  {\frontmatter@RRAPformat{\produce@RRAP{*#1\href{mailto:#2}{#2}}}\frontmatter@RRAPformat}
  {}{}
}%
\makeatother
\begin{document}

\preprint{AIP/123-QED}

\title[preprint]{A symbolic regression-based implicit algebraic stress turbulence model: incorporating the production of non-dimensional Reynolds stress deviatoric tensor}
\author{Ziqi Ji 
}
\affiliation{%
School of Energy and Power Engineering, Beihang University, Beijing, 100190, China.%
}%
\affiliation{%
Department of Mechanical Engineering, City University of Hong Kong, Hong Kong, 999077, China.%
}%

\author{Penghao Duan 
\thanks}%
\affiliation{%
Department of Mechanical Engineering, City University of Hong Kong, Hong Kong, 999077, China.%
}%

\author{Gang Du 
\thanks}
\affiliation{%
School of Energy and Power Engineering, Beihang University, Beijing, 100190, China.%
}%

\date{\today}

\begin{abstract}
Turbulence constitutes an exceptionally complex and irregular flow phenomenon that manifests in liquids, gases, and plasma, making it ubiquitous in both natural processes and engineering applications. Given the relatively modest advancements in classical turbulence models over the past half-century, data-driven approaches, such as machine learning, have recently gained considerable traction in turbulence model research. In this study, we introduce a symbolic regression-based implicit algebraic stress turbulence model that incorporates the production of non-dimensional Reynolds stress deviatoric tensor, thereby capturing the contribution of the shape of local turbulence produced by the mean flow field. We rigorously evaluate our model across five distinct characteristic flow cases and benchmark it against three alternative turbulence models. Our comprehensive analysis demonstrates that the proposed model exhibits robust performance and substantial generalizability across all test cases while manifesting notable advantages when compared with the reference turbulence models.
\end{abstract}

\maketitle

\section{\label{sec:Introduction}Introduction}

Turbulence represents an extraordinarily complex and irregular flow phenomenon occurring in liquids, gases, and plasma, rendering it omnipresent in both natural processes and engineering applications \cite{pope2001turbulent}. Nevertheless, conducting expeditious and high-fidelity Computational Fluid Dynamics (CFD) simulations for high Reynolds number flows remains challenging due to the prohibitive computational demands of Large Eddy Simulation (LES) and Direct Numerical Simulation (DNS), coupled with the inherent limitations in accuracy exhibited by conventional turbulence models in Reynolds-averaged Navier--Stokes (RANS) simulations. The progressive advancement of machine learning technologies has facilitated their increasing integration into turbulence modeling frameworks. Consequently, numerous machine learning-based turbulence models have emerged, demonstrating remarkable efficacy and yielding exceptional results.

The uncertainties introduced by RANS turbulence models can be categorized into four physical levels \citep{duraisamy_turbulence_2019,he_turbo-oriented_2022}:

Physical level 1: The loss of fluctuation information due to the ensemble averaging of the RANS equations, which is an inevitable consequence of the RANS method.

Physical level 2: The uncertainty arising from the assumptions regarding the constitutive relationship between the Reynolds stress tensor and the mean flow quantities (for instance, the uncertainties brought about by the Boussinesq hypothesis).

Physical level 3: The uncertainty originating from the form of the turbulence model transport equation.

Physical level 4: The uncertainty associated with the calibration of the turbulence model coefficients.

Physical level 2 represents the most significant uncertainty physical level in RANS turbulence models and has attracted substantial attention in machine learning turbulence modeling research. The majority of machine learning turbulence model studies focusing on physical level 2 are founded upon the general effective-viscosity hypothesis proposed by Pope \cite{pope_more_1975}, which postulates that the non-dimensional Reynolds stress deviatoric tensor can be expressed as a finite tensor polynomial.

Nevertheless, despite the general effective-viscosity model (GEVM) representing a more comprehensive extension of the Boussinesq hypothesis and having been extensively implemented in various machine learning-based turbulence models \cite{ling_reynolds_2016, wu_priori_2017, wu_physics-informed_2018, yin_feature_2020, strofer_end--end_2021, zhang_ensemble_2022, xu_pde-free_2022, liu_learning_2023, zhang_physical_2023, zhang_combining_2023, amarloo_data-driven_2023, sun_development_2024, liu_data_2024, ji_tensor_2024, shan_new_2025, chenyu_field_2025, zafar_data-driven_2025, weatheritt_novel_2016, zhao_rans_2020, schmelzer_discovery_2020, ben_hassan_saidi_cfd-driven_2022, xie_data-driven_2023, fang_toward_2023, lav_coupled_2023, wu_enhancing_2023, he_field_2024, wu_development_2024, li_evolutionary_2024, fang2024exploiting, ji_tensor_2024, stocker_dns-based_2024, ji_interpreting_2025, ji2025enhancinggeneralizabilitymachinelearning, shan_improved_2025, sun2025generalizedsamodelsymbolic, shan_data-driven_2025, zhang_numerical_nodate, cherroud_space-dependent_2025, oulghelou2025machine}, it fundamentally fails to account for Reynolds stress transport phenomena. Consequently, researchers face substantial challenges in establishing clear physical correspondences between Reynolds stress transport equations and the GEVM framework.

The algebraic stress model (ASM), introduced by Rodi \cite{rodi_new_1976} in 1976, operates on a well-established physical premise: that Reynolds stress transport is proportional to turbulence kinetic energy transport. This formulation establishes a rigorous mathematical relationship with the Reynolds stress transport equation, thereby offering a more physically coherent and theoretically sound framework compared to the GEVM.

However, due to limitations in robustness, the ASM has not gained widespread adoption since its initial proposal was made. The recent emergence of machine learning techniques has revolutionized the field of turbulence modeling. Among these approaches, symbolic regression has demonstrated exceptional generalizability and interpretability \cite{weatheritt_novel_2016, zhao_rans_2020, schmelzer_discovery_2020, ben_hassan_saidi_cfd-driven_2022, xie_data-driven_2023, fang_toward_2023, lav_coupled_2023, wu_enhancing_2023, he_field_2024, wu_development_2024, li_evolutionary_2024, fang2024exploiting, ji_tensor_2024, stocker_dns-based_2024, ji_interpreting_2025, ji2025enhancinggeneralizabilitymachinelearning, shan_improved_2025, sun2025generalizedsamodelsymbolic, shan_data-driven_2025, zhang_numerical_nodate, cherroud_space-dependent_2025, oulghelou2025machine}. These suggest promising potential for developing a symbolic regression-based ASM framework.

Generalizability represents a critical attribute for machine learning-based turbulence models. In their seminal work, Wu et al. \cite{wu_development_2024} introduced a systematic and quantitative evaluation framework in 2024 for assessing the generalization capabilities of turbulence models (as presented in Table \ref{Table: Generalizaion_level_of_1_to_4}). Subsequently, numerous methodological approaches \cite{wu_development_2024, bin_constrained_2024, ji2025enhancinggeneralizabilitymachinelearning} have been proposed to enhance the generalization performance of machine learning-based turbulence models across diverse flow configurations.

\begin{table*}[htbp]
\caption{Generalizaion level of 1 to 4 \citep{wu_development_2024}}
\begin{ruledtabular}
\begin{tabular}{p{0.2\linewidth} p{0.25\linewidth} p{0.45\linewidth}}
\textbf{Level} & \textbf{Definition} & \textbf{Example} \\
\hline
Generalization level 1 & The model performs well in a series of geometries similar to the training set. & The model is trained on one periodic hill and can be generalized to other periodic hills with different aspect ratios. \\ 
Generalization level 2 & The model does not negatively affect the baseline model's accuracy in simple wall-attached flows. & The model is trained on some separated flow cases and can be as accurate as its baseline model on the zero-pressure-gradient flat plate. \\ 
Generalization level 3 & The model performs well in test cases that have separation features similar to the training set but completely different geometries and Reynolds numbers. & (1) The model is trained on a case where separation is caused by a blunt geometry and is tested to be effective in other cases that have completely different blunt geometries and Reynolds numbers. 
        
(2) The model is trained on a case where separation starts from a smooth surface (caused by a negative pressure gradient) and is tested to be effective on other smooth surface separations that have completely different geometries and Reynolds numbers. \\
Generalization level 4 & The model performs well in a series of test cases with completely different separation features, geometries, and Reynolds numbers. & The model can accurately predict separation caused by blunt geometry as well as separation starting from a smooth surface. \\ 
\end{tabular}
\end{ruledtabular}
\label{Table: Generalizaion_level_of_1_to_4}
\end{table*}

The remainder of this paper is structured as follows: Sec. \ref{sec:Methodology} outlines the methodological framework employed in this investigation. Sec. \ref{sec:Results} presents the CFD simulation results and comparative analyses. Sec. \ref{sec: Discussion} examines the fundamental physical advantages of our symbolic regression-based implicit algebraic stress turbulence model compared to explicit models, while also addressing the pivotal role of machine learning techniques in turbulence modeling. Finally, Sec. \ref{sec: Conclusion} summarizes the key conclusions of this work.

\section{Methodology}\label{sec:Methodology}

\subsection{\label{sec:The origin formulation of Rodi's algebraic stress mode}The origin formulation of algebraic stress model}

The transport equation for the Reynolds stress tensor $\overline{u'_i u'_j}$ can be expressed as\cite{rodi_new_1976}:
\begin{equation}
\begin{aligned}
\underbrace{\frac{D \overline{u'_i u'_j}}{D t}}_{\text { convection }}=  \underbrace{c_s \frac{\partial}{\partial x_k}\left(\frac{k}{\varepsilon} \overline{u'_k u'_l} \frac{\partial \overline{u'_i u'_j}}{\partial x_l}\right)}_{D_{i j}=\text { diffusion }} \underbrace{-\overline{u'_i u'_k} \frac{\partial U_i}{\partial x_k}-\overline{u'_j u'_k} \frac{\partial U_i}{\partial x_k}}_{P_{i j}=\text { production }} \\
 \underbrace{-c_1 \frac{\varepsilon}{k}\left(\overline{u'_i u'_j}-\delta_{i j} \frac{2}{3} k\right)-\gamma\left(P_{i j}-\delta_{i j} \frac{2}{3} P\right)}_{\text {pressure strain }}\underbrace{-\frac{2}{3} \delta_{i j} \varepsilon}_\text { dissipation },
\end{aligned}
\label{eq:transport equation for Reynolds stress}
\end{equation}
where $u'$ denotes the fluctuation velocity, $U$ represents the mean velocity. The term $P$ signifies the production of turbulence, $\varepsilon$ corresponds to the dissipation rate of turbulent kinetic energy $k$, and $c_s$, $c_1$, and $\gamma$ are empirically determined constants. The detailed formulation of the production term $P$ is presented in the transport equation for turbulent kinetic energy $k$ as follows:
\begin{equation}
\frac{D k}{D t}=\underbrace{c_s \frac{\partial}{\partial x_k}\left(\frac{k}{\varepsilon} \overline{u'_k u'_l} \frac{\partial k}{\partial x_l}\right)}_{D_k} \underbrace{-\overline{u'_k u'_l} \frac{\partial U_k}{\partial x_l}}_{P=\frac{1}{2} P_{i i}}-\varepsilon .
\label{eq:transport equation for turbulence kinetic energy}
\end{equation}

In 1976, Rodi \cite{rodi_new_1976} proposed a modeling approach that postulates that the transport of Reynolds stress is proportional to the transport of turbulent kinetic energy, with the proportionality coefficient defined as $\frac{\overline{u'_i u'_j}}{k}$:
\begin{equation}
\frac{D \overline{u'_i u'_j}}{D t}-D_{i j}=\frac{\overline{u'_i u'_j}}{k}\left(\frac{D k}{D t}-D_k\right)=\frac{\overline{u'_i u'_j}}{k}(P-\varepsilon) .
\label{eq:modeling}
\end{equation}

By incorporating Eq. (\ref{eq:modeling}) into Eq. (\ref{eq:transport equation for Reynolds stress}), we can derive the algebraic stress model initially proposed by Rodi \cite{rodi_new_1976}:
\begin{equation}
\overline{u_i u_j}=k\left[\frac{2}{3} \delta_{i j}+\frac{1-\gamma}{c_1} \frac{P_{i j} / \varepsilon-\frac{2}{3} \delta_{i j} P / \varepsilon}{1+\frac{1}{c_1}(P / \varepsilon-1)}\right]
\label{eq:ASM}
\end{equation}

However, the original version of ASM was not widely adopted due to its lack of robustness. Therefore, this paper aims to enhance the original ASM framework by implementing the symbolic regression methodology.

\subsection{\label{sec:Non-dimensional formulation of the algebraic stress model}Non-dimensional formulation of the algebraic stress model}

When constructing machine learning-based turbulence models, non-dimensional data significantly facilitates algorithm processing and implementation. Therefore, it is necessary to derive a dimensionless formulation of Eq. (\ref{eq:ASM}).

The production term of the Reynolds stress tensor, $P_{ij}$, can be organized as follows: 
\begin{equation}
\begin{aligned}
\frac{P_{i j}}{2 k}= &-\left(\overline{\frac{u_i^{\prime} u_k^{\prime}}{2 k}}-\frac{1}{3} \delta_{i k}\right) \frac{\partial U_j}{\partial x_k}-\frac{1}{3} \delta_{i k} \frac{\partial U_j}{\partial x_k} \\ 
& -\left(\frac{\overline{u_j^{\prime} u_k^{\prime}}}{2 k}-\frac{1}{3} \delta_{j k}\right) \frac{\partial U_i}{\partial x_k}-\frac{1}{3} \delta_{j k} \frac{\partial U_i}{\partial x_k} \\ =  & -b_{i k} \frac{\partial U_j}{\partial x_k}-\frac{1}{3} \delta_{i k} \frac{\partial U_j}{\partial x_k}-b_{j k} \frac{\partial U_i}{\partial x_k}-\frac{1}{3} \delta_{j k} \frac{\partial U_i}{\partial x_k},
\end{aligned}
\label{eq:non-dimensional Pij}
\end{equation}
where $b_{ij}=\overline{u_i^{\prime} u_j^{\prime}} / 2k-\frac{1}{3}\delta_{i j}$ is the non-dimensional Reynolds stress deviatoric tensor.

The production term of the turbulent kinetic energy, $P$, can be organized as follows:
\begin{equation}
\begin{aligned}
 \frac{P}{2 k}&=-\left(\frac{\overline{u_i^{\prime} u_j^{\prime}}}{2 k}-\frac{1}{3} \delta_{i j}\right) \frac{\partial U_i}{\partial x_j}-\frac{1}{3} \delta_{i j} \frac{\partial U_i}{\partial x_j} \\
& =-b_{i j} \frac{\partial U_i}{\partial x_j}-\frac{1}{3} \delta_{i j} \frac{\partial U_i}{\partial x_j}.
\end{aligned}
\label{eq:non-dimensional P}
\end{equation}

To enhance the readability of Eqs. (\ref{eq:non-dimensional Pij}) and (\ref{eq:non-dimensional P}), and to facilitate a more straightforward derivation process, we reformulate these equations in their tensorial form as follows:
\begin{equation}
\begin{aligned}
\left\{
\begin{aligned}
\frac{\mathbf{P}}{2 k} &=-\left(\mathbf{b}(\nabla \mathbf{U})^T+(\nabla \mathbf{U}) \mathbf{b}^T\right)-\frac{2}{3} \mathbf{S}, \\
\frac{P}{2 k} &=-\mathbf{b}:(\nabla \mathbf{U})-\frac{1}{3} \operatorname{Tr}(\nabla \mathbf{U}),
\end{aligned}
\right.
\end{aligned}
\label{eq:tensor version}
\end{equation}
where $\mathbf{S}=\frac{1}{2}\left(\nabla \mathbf{U}+(\nabla \mathbf{U})^{\mathrm{T}}\right)$ denotes the mean strain-rate tensor, ``T'' signifies the transpose, and ``:'' represents the double dot product. ``Tr($\cdot$)'' denotes the trace of a tensor.

We define a non-dimensional mean velocity gradient as $\nabla \widehat{\mathbf{U}}=\frac{k}{\varepsilon} \nabla \mathbf{U}$ and a non-dimensional mean strain-rate tensor as $\widehat{\mathbf{S}}=\frac{k}{\varepsilon} \mathbf{S}$. Consequently, Eq. (\ref{eq:tensor version}) can be expressed in a non-dimensional form as follows:
\begin{equation}
\left\{
\begin{aligned}
\frac{k}{\varepsilon}\frac{\mathbf{P}}{2 k} &=-\left(\mathbf{b}(\nabla \widehat{\mathbf{U}})^T+(\nabla \widehat{\mathbf{U}}) \mathbf{b}^T\right)-\frac{2}{3} \widehat{\mathbf{S}} \\
\frac{k}{\varepsilon}\frac{P}{2 k} &=-\mathbf{b}:(\nabla \widehat{\mathbf{U}})-\frac{1}{3} \operatorname{Tr}(\nabla \widehat{\mathbf{U}})
\end{aligned}
\right..
\label{eq:non-dimensional tensor version}
\end{equation}

Substituting Eq. (\ref{eq:non-dimensional tensor version}) into Eq. (\ref{eq:ASM}), we obtain the tensor formulation of the non-dimensional Reynolds stress deviatoric tensor $\mathrm{b}$, which serves as the foundation for the subsequent analysis in this study:
\begin{widetext}
\begin{equation}
\begin{aligned}
\mathbf{b} &=\frac{1-\gamma}{\frac{1}{2}\left(C_1-1\right)+\frac{k}{\varepsilon} \frac{P}{2 k}}\left(\frac{1}{2} \frac{k}{\varepsilon} \frac{\mathbf{P}}{2 k}-\frac{1}{3} \frac{k}{\varepsilon} \frac{P}{2 k} \mathbf{I}\right) \\
&=\underbrace{\frac{-\frac{1}{2}\left(1-\gamma\right)}{\frac{1}{2}\left(C_1-1\right)-\mathbf{b}:(\nabla \widehat {\mathbf{U}})-\frac{1}{3} \operatorname{Tr}(\nabla \widehat {\mathbf{U}})}}_{\text {Scalar }} \underbrace{\left\{\left[\left(\mathbf{b}(\nabla \widehat{\mathbf{U}})^T+(\nabla \widehat{\mathbf{U}}) \mathbf{b}^T\right)-\frac{2}{3} \mathbf{b}:(\nabla \widehat{\mathbf{U}}) \mathbf{I}\right]+\frac{2}{3}\left[\widehat{\mathbf{S}}-\frac{1}{3} \operatorname{Tr}(\widehat{\mathbf{S}}) \mathbf{I}\right]\right\}}_{\text {Tensor }},
\end{aligned}
\label{eq:non-dimensional tensor version b}
\end{equation}
\end{widetext}
where $\mathbf{I}$ denotes the third-order identity matrix.

\subsection{\label{sec:Machine learning formulation of the algebraic stress model}Machine learning formulation of the algebraic stress model}

To enhance the efficiency of the machine learning algorithm, we employ the local normalization method \cite{wu_physics-informed_2018}, which is commonly used in machine learning-based turbulence models:
\begin{equation}
\hat{\mathbf{\alpha}}=\frac{\mathbf{\alpha}}{\|\mathbf{\alpha}\|+|\beta|}.
\label{eq: local normalization scheme}
\end{equation}
Consequently, the mean velocity gradient and mean strain-rate tensor can be normalized by Eq. (\ref{eq: local normalization scheme}) as follows:
\begin{equation}
\begin{aligned}
\left\{
\begin{aligned}
\nabla \hat{\mathbf{U}}&=\frac{\nabla \mathbf{U}}{||\nabla \mathbf{U}||_F+\omega}\\\hat{\mathbf{S}}&=\frac{\mathbf{S}}{||\mathbf{S}||_F+\omega},
\end{aligned}
\right.
\end{aligned}
\label{eq: gradU_S}
\end{equation}
where $\omega = \frac{\varepsilon}{0.09k}$ denotes the specific dissipation rate. Here, $||\cdot||_F$ represents the Frobenius norm. It is important to note that this normalization approach differs from the methodology employed in Section \ref{sec:Non-dimensional formulation of the algebraic stress model}.

Therefore, we can define a tensor $\mathbf{T}$ as follows:
\begin{widetext}
\begin{equation}
\begin{aligned}
\mathbf{T} \triangleq \left[\left(\mathbf{b}(\nabla \hat{\mathbf{U}})^T+(\nabla \hat{\mathbf{U}}) \mathbf{b}^T\right)-\frac{2}{3} \mathbf{b}:(\nabla \hat{\mathbf{U}}) \mathbf{I}\right]+\frac{2}{3}\left[\hat{\mathbf{S}}-\frac{1}{3} \operatorname{Tr}(\hat{\mathbf{S}}) \mathbf{I}\right]
\end{aligned}
\label{eq: T}
\end{equation}
\end{widetext}

Based on Eqs. (\ref{eq:non-dimensional tensor version b}) and (\ref{eq: T}), we can postulate that the non-dimensional Reynolds stress deviatoric tensor $\mathbf{b}$ be expressed as:
\begin{equation}
\begin{aligned}
\mathbf{b} = g \mathbf{T},
\end{aligned}
\label{eq:machine learning ASM}
\end{equation}
where $g$ denotes an undetermined scalar coefficient that can be optimized through machine learning techniques, by Eq. (\ref{eq:machine learning ASM}), $g$, which serves as the target label in the machine learning framework, can be calculated as follows:
\begin{equation}
g=\underset{g}{\arg \min }\left(\left\|\mathbf{b}-g \mathbf{T}\right\|_F^2\right)=\frac{\mathbf{b}: \mathbf{T}}{\left\|\mathbf{T}\right\|_F^2},
\label{eq: g}
\end{equation}

To ensure that the prediction target value of the machine learning algorithm remains within a reasonable range, we employ the tensor basis normalization method \cite{ji2025enhancinggeneralizabilitymachinelearning}. This approach defines a normalized tensor $\hat{\mathbf{T}}\triangleq\frac{\mathbf{T}}{\left||\mathbf{T}\right||_F}$ and a corresponding normalized coefficient $\hat{g}\triangleq \left||\mathbf{T}\right||_F g$. Consequently, Eq. (\ref{eq:machine learning ASM}) can be reformulated as:
\begin{equation}
\mathbf{b}= \hat{g} \hat{\mathbf{T}}.
\label{eq: machine learning ASM ghat}
\end{equation}
Consequently, the normalized coefficient $\hat{g}$, which serves as the target label in the machine learning framework, can be calculated through the following expression:
\begin{align}
\hat{g}=\mathbf{b}: \frac{\mathbf{T}}{\left\|\mathbf{T}\right\|_F}.
\label{eq: g_normalized}
\end{align}

Due to the structural similarity between Eq. (\ref{eq: machine learning ASM ghat}) and established machine learning-based general effective-viscosity turbulence models \cite{ling_evaluation_2015, weatheritt_novel_2016}, we can adopt the same physical assumptions regarding the normalized coefficient $\hat{g}$:
\begin{equation}
\hat{g}=f\left(I_1 \sim I_a, q_1 \sim q_b, ||\mathbf{T}||_F\right),
\label{eq: gi_assumpsion}
\end{equation}
where $\left(I_1, \ldots, I_a\right)$ represent the tensor invariants, while $\left(q_1, \ldots, q_b\right)$ denote additional mean flow characteristics. Since the tensor basis normalization process eliminates norm information, we incorporate the Frobenius norm $||\mathbf{T}||_F$ into the function. The coefficients $\hat{g}$ are functions of both the tensor invariants $\left(I_1, \ldots, I_a\right)$ and the aforementioned mean flow characteristics $\left(q_1, \ldots, q_b\right)$.

\subsection{\label{sec:Selection of tensor invariants and extra mean flow characteristics}Selection of tensor invariants and extra mean flow characteristics}

The tensor invariants employed in this study were originally introduced by Wu et al. \cite{wu_physics-informed_2018}. However, due to computational constraints, rather than utilizing all 47 tensor invariants, we strategically selected 7 specific invariants based on their order, following the approach proposed in our previous research \cite{ji_tensor_2024}. To maintain consistency with prior studies, we preserved the original indexing system established by Wu et al. \cite{wu_physics-informed_2018}. Table \ref{Table: Invariant basis} presents these 7 selected tensor invariants. All tensors listed in Table \ref{Table: Invariant basis} have been normalized according to Eq. (\ref{eq: local normalization scheme}), with the corresponding normalization factors detailed in Table \ref{Table: input normalization}.

\begin{table}[htbp]
\caption{Tensor invariants are the trace of these selected tensors.}
\begin{ruledtabular}
\begin{tabular}{cc}
\textbf{Feature Index} & \textbf{Invariant Basis} \\
\hline
1 & $\hat{\mathbf{S}}^2$ \\ 
3-5 & $\hat{\mathbf{R}}^2, \hat{\mathbf{A}}_p^2, \hat{\mathbf{A}}_k^2$ \\ 
15-17 & $\hat{\mathbf{R}} \hat{\mathbf{A}}_p, \hat{\mathbf{A}}_p \hat{\mathbf{A}}_k, \hat{\mathbf{R}} \hat{\mathbf{A}}_k$ \\ 
\end{tabular}
\end{ruledtabular}
\label{Table: Invariant basis}
\end{table}

\begin{table*}[htbp]
\caption{\label{Table: input normalization}The normalization of mean strain rate tensor, mean rotation rate tensor, pressure gradient, and turbulence kinetic energy gradient.}
\begin{ruledtabular}
\begin{tabular}{cccc}
Normalized raw input $\hat{\mathbf{\alpha}}$ & Description & Raw input $\mathbf{\alpha}$ & Normalization factor $\beta$ \\ \hline
$\nabla \hat{\mathbf{U}}$ & Mean velocity gradient tensor & $\nabla {\mathbf{U}}$ & $\omega$ \\
$\hat{\mathbf{S}}$ & Mean strain rate tensor & $\mathbf{S}$ & $\omega$ \\
$\hat{\mathbf{R}}$ & Mean rotation rate tensor & $\mathbf{R}$ & $\omega$ \\
$\hat{\nabla p}$ & Pressure gradient & $\nabla p$ & $\omega \sqrt{k}$ \\
$\hat{\nabla k}$ & Turbulence kinetic energy gradient & $\nabla k$ & $\omega \sqrt{k}$ \\
\end{tabular}
\end{ruledtabular}
\end{table*}

We incorporate some extra features outlined in Table \ref{Table: q} to enhance prediction accuracy.  $d$ denotes the distance to the nearest wall, $\nu$ is the kinematic viscosity, $\varepsilon$ is the turbulence dissipation rate, and $c$ stands for the local speed of sound. In this work, the turbulence specific dissipation rate $\omega$ is computed as follows:
\begin{equation}
\omega = \frac{\varepsilon}{0.09k}.
\label{eq: ep}
\end{equation}

Moreover, the following formula was employed for non-dimensionalization:
\begin{equation}
q_\beta=\frac{\hat{q}_\beta}{\left|\hat{q}_\beta\right|+\left|q_\beta^*\right|},
\label{eq: extra features non-dimensionalization}
\end{equation}
except for $q_2$ and $q_7$, the wall-distance based Reynolds number and Production of non-dimensional Reynolds stress deviatoric tensor are inherently dimensionless and therefore do not require non-dimensionalization.

\begin{table*}[htbp]
\caption{\label{Table: q}List of extra features.}
\begin{ruledtabular}
\begin{tabular}{>{\centering\arraybackslash}p{30mm}>{\centering\arraybackslash}p{35mm}>{\centering\arraybackslash}p{30mm}>{\centering\arraybackslash}p{25mm}}
Normalized extra features $q_\beta$ & Description & Origin extra features $\hat{q}_\beta$ & Normalization factor $q_\beta^*$ \\ \hline
$q_1$ & Ratio of excess mean rotation rate to mean strain rate (Q criterion) & $\frac{1}{2}\left(\|\mathbf{R}\|_F^2-\|\mathbf{S}\|_F^2\right)$ & $\|\mathbf{S}\|_F^2$ \\ 
$q_2$ & Wall-distance based Reynolds number & $\min \left(\frac{\sqrt{k} \cdot d}{50 \nu}, 2\right)$ & Not applicable \\ 
$q_3$ & Ratio of turbulent time scale to mean strain time scale & $\frac{k}{\varepsilon}$ & $\frac{1}{||\mathbf{S}||_F}$ \\ 
$q_4$ & Turbulent kinetic energy & $k$ & $\nu ||\mathbf{S}||_F$ \\ 
$q_{5}$ & Ratio of turbulent time scale to mean strain time scale & $\|\mathbf{S}\|_F$ & $\omega$  \\
$q_{6}$ & Ratio of turbulent/mean viscosity & $\nu_t$ & $\nu$ \\
$q_{7}$ & Production of non-dimensional Reynolds stress deviatoric tensor & $\mathbf{b}:(\nabla \hat{\mathbf{U}})$ & Not applicable \\
\end{tabular}
\end{ruledtabular}
\end{table*}

\subsection{\label{sec: Symbolic regression}Symbolic regression}

This work uses open-source multi-population evolutionary symbolic regression library PySR \cite{cranmer2023interpretablemachinelearningscience}.

Within our framework, the symbolic regression is trained to predict normalized coefficients $\hat{g}$. The regression input comprises tensor invariants, extra mean flow features, and the Frobenius norm of the tensor $||\mathbf{T}||_F$, as elaborated in Sec. \ref{sec:Machine learning formulation of the algebraic stress model} and \ref{sec:Selection of tensor invariants and extra mean flow characteristics}.

The key hyperparameters of our symbolic regression algorithm are enumerated in Table \ref{Table: Main hyperparameters of the algorithm}. Our implementation incorporates 8 distinct operators as detailed in Table \ref{Table: Main hyperparameters of the algorithm}. The "neg" operator performs negation, "inv" computes the inverse function, while "square" and "cubic" execute second and third power operations, respectively. Each constant within the formulation is assigned a complexity value of 2. The algorithmic framework employs 16 populations, each containing 100 individuals. During each iteration, we perform 500 mutations for every ten samples extracted from each population, continuing this process for 100 iterations. The initialization phase comprises 200 iterations. We impose a maximum complexity constraint of 20 for any generated equation, with a maximum depth threshold of 10. Additional hyperparameters conform to the default configurations specified in the PySR library. Furthermore, we implement the symbolic regression strategy introduced in our previous work \cite{ji2025enhancinggeneralizabilitymachinelearning}, which ensures convergence of the results. The symbolic regression training outcome obtained through this strategy is as follows:
\begin{equation}
\hat{g}=-0.292\left(q_3+I_5\right)
\label{eq: g_hat}
\end{equation}

\begin{table*}[htbp]
\caption{\label{Table: Main hyperparameters of the algorithm}Main hyperparameters of the symbolic regression algorithm.}
\begin{ruledtabular}
\begin{tabular}{cc}
Principal parameters of the algorithm & Parameters set-up \\ \hline
Operators & +, -, $\times$, $\div$, neg, inv, square, cubic \\ 
Complexity of constants & 2 \\ 
Number of selected features & 10 \\ 
Number of populations & 8 \\ 
Number of individuals in each population & 100 \\ 
Number of total mutations to run, per 10 samples of the population, per iteration & 500 \\ 
Number of beginning iterations & 200 \\ 
Max complexity of an equation & 20 \\ 
Max depth of an equation & 10 \\ 
\end{tabular}
\end{ruledtabular}
\end{table*}

\subsection{Baseline model and other models}
\label{sec:Baseline model and other models}

\subsubsection{Baseline model}
\label{sec:Baseline model}

The baseline model we used in this work is $k$-$\epsilon$ model proposed by Launder and Spalding \cite{launder1974application}. The governing equation of baseline model can be expressed by:
\begin{widetext}
\begin{align}
\left\{\begin{aligned}
\frac{\partial \rho k}{\partial t}+\nabla \cdot(\rho \mathbf{U} k)-\nabla \cdot\left(\rho D_k \nabla k\right) &= \rho G-\frac{2}{3} \rho(\nabla \cdot \mathbf{U}) k-\rho \frac{\varepsilon+D}{k} k+S_k \\
\frac{\partial \rho \varepsilon}{\partial t}+\nabla \cdot(\rho \mathbf{U} \varepsilon)-\nabla \cdot\left(\rho D_{\varepsilon} \nabla \varepsilon\right) &= C_1 \rho G \frac{\varepsilon}{k}-\left(\left(\frac{2}{3} C_1-C_3\right) \rho(\nabla \cdot \mathbf{U}) \varepsilon\right) -C_2 f_2 \rho \frac{\varepsilon}{k} \varepsilon+\rho E+S_{\varepsilon}
\end{aligned}\right.
\label{eq: k-epsilon}
\end{align}
\end{widetext}
where $G=\nu_t\left(\nabla \mathbf{U}+\nabla \mathbf{U}^T-\tfrac{2}{3}(\nabla \cdot \mathbf{U}) \mathbf{I}\right): \nabla \mathbf{U}$ represents the turbulence production term, $E=2 \nu \nu_t|\nabla \nabla \mathbf{U}|^2$ denotes the wall-reflection term, and $f_2=1-0.3 \exp \left(-\min \left(\smash{\tfrac{k^4}{\nu^2 \varepsilon^2}}, 50\right)\right)$ is a damping function. Additionally, $D=2 \nu|\nabla(\sqrt{k})|^2$ signifies the diffusion term, $f_\mu=\exp \left({-3.4 / \left(1+{k^2 \over 50 \nu \varepsilon}\right)^2}\right)$ is another damping function, and $\nu_t=C_\mu f_\mu \tfrac{k^2}{\varepsilon}$ represents the eddy viscosity. The diffusion coefficients are given by $D_k=\tfrac{\nu_t}{\sigma_k}+\nu$ and $D_{\varepsilon}=\tfrac{\nu_t}{\sigma_{\varepsilon}}+\nu$ for the $k$ and $\varepsilon$ equations, respectively. Other empirical coefficients can be referenced in Launder and Spalding \cite{launder1974application}.

\subsubsection{General effective-viscosity model}
\label{sec:General effective-viscosity model}

The GEVM was initially introduced by Pope \cite{pope_more_1975} in 1975. In the present study, we employ the normalized tensor basis formulation \cite{ji2025enhancinggeneralizabilitymachinelearning} of the quadratic GEVM:
\begin{align}
\left\{\begin{array}{l}
\mathbf{b}=\sum_{i=1}^4 \hat{g}_i \hat{\mathbf{T}}_i \\
\hat{g}_i=\hat{f}\left(I_1 \sim I_a, q_1 \sim q_b\right).
\end{array}\right.
\label{eq: gihat}
\end{align}

The normalized coefficients $\hat{g}_i$ results training by using the input features in Sec. \ref{sec:Selection of tensor invariants and extra mean flow characteristics} and the symbolic regression algorithm in Sec. \ref{sec: Symbolic regression} are shown below:
\begin{align}
\left\{
\begin{array}{l}
\hat{g}_1 = -0.0721\left[\left(q_3+I_4\right) q_2\right]^2 \\
\hat{g}_2 = q_4^{18}-1.14 \\
\hat{g}_3 = -0.431 I_{16}^2 \\
\hat{g}_4 = 0.434 I_{16}^2
\end{array}
\right.
\label{eq: gihat_SR}
\end{align}

\subsubsection{A turbulence model constructed using the kinetic Fokker–Planck equation}
\label{sec:A turbulence model constructed using the kinetic Fokker–Planck equation}

For comparative analysis, we employ a turbulence model derived from the kinetic Fokker-Planck equation. This model was constructed based on the analogy between the Brownian motion of molecules and the dynamics of turbulent flow. As proposed by Luan et al. \cite{luan_constructing_2025}, the model is governed by the following equation:
\begin{equation}
\begin{aligned}
\mathbf{b} = &-\frac{v_t}{k}\left[\mathbf{S}-\frac{1}{3} \operatorname{tr}(\mathbf{S}) \mathbf{I}\right] +\frac{v_t^2}{2 k^2 / 3}\left[\mathbf{U} \cdot \nabla\left(\mathbf{S}-\frac{1}{3} \operatorname{tr}(\mathbf{S}) \mathbf{I}\right)\right]\\ 
&+\frac{2 v_t^2}{2 k^2 / 3}\left[\mathbf{S}^2-\frac{1}{3} \operatorname{tr}\left(\mathbf{S}^2\right) \mathbf{I}\right]-\frac{v_t^2}{2 k^2 / 3}(\mathbf{S} \mathbf{R}-\mathbf{R} \mathbf{S}).
\end{aligned}
\label{eq: KFP}
\end{equation}
This turbulence model is referred to as ``KFPM'' throughout this study.

\subsection{Framework}
\label{sec:Framework}

In this study, we employ TCAE as our CFD software for conducting simulations. TCAE, developed as an extension of the open-source CFD platform OpenFOAM, features a proprietary solver called ``redSolver'' that demonstrates exceptional robustness when modeling compressible and transonic flow regimes. In our investigation, this solver is specifically utilized to simulate transonic axial compressor rotors.

We employ certain turbulent quantities within the proposed framework, specifically the turbulence kinetic energy $k$ and the specific dissipation rate $\omega$, for normalization purposes, as delineated in Table \ref{Table: input normalization}. Due to discrepancies between the statistical turbulent quantities computed by DNS and the modeled turbulent quantities derived from RANS, it becomes important to utilize turbulent quantities consistent with the high-fidelity mean flow fields as a normalization factor. We commence by interpolating high-fidelity mean flow data onto the RANS computational grid to acquire compatible turbulent quantities. Subsequently, the interpolated high-fidelity mean flow data are introduced into the $k$ equation and $\omega$ equation. The mean flow is held constant throughout this process, yielding a compatible turbulence flow field, as outlined in the iterative methodology described by Yin et al. \cite{yin_iterative_2022}.

\begin{figure*}[htbp]
\centering \includegraphics[width=1.0\textwidth]{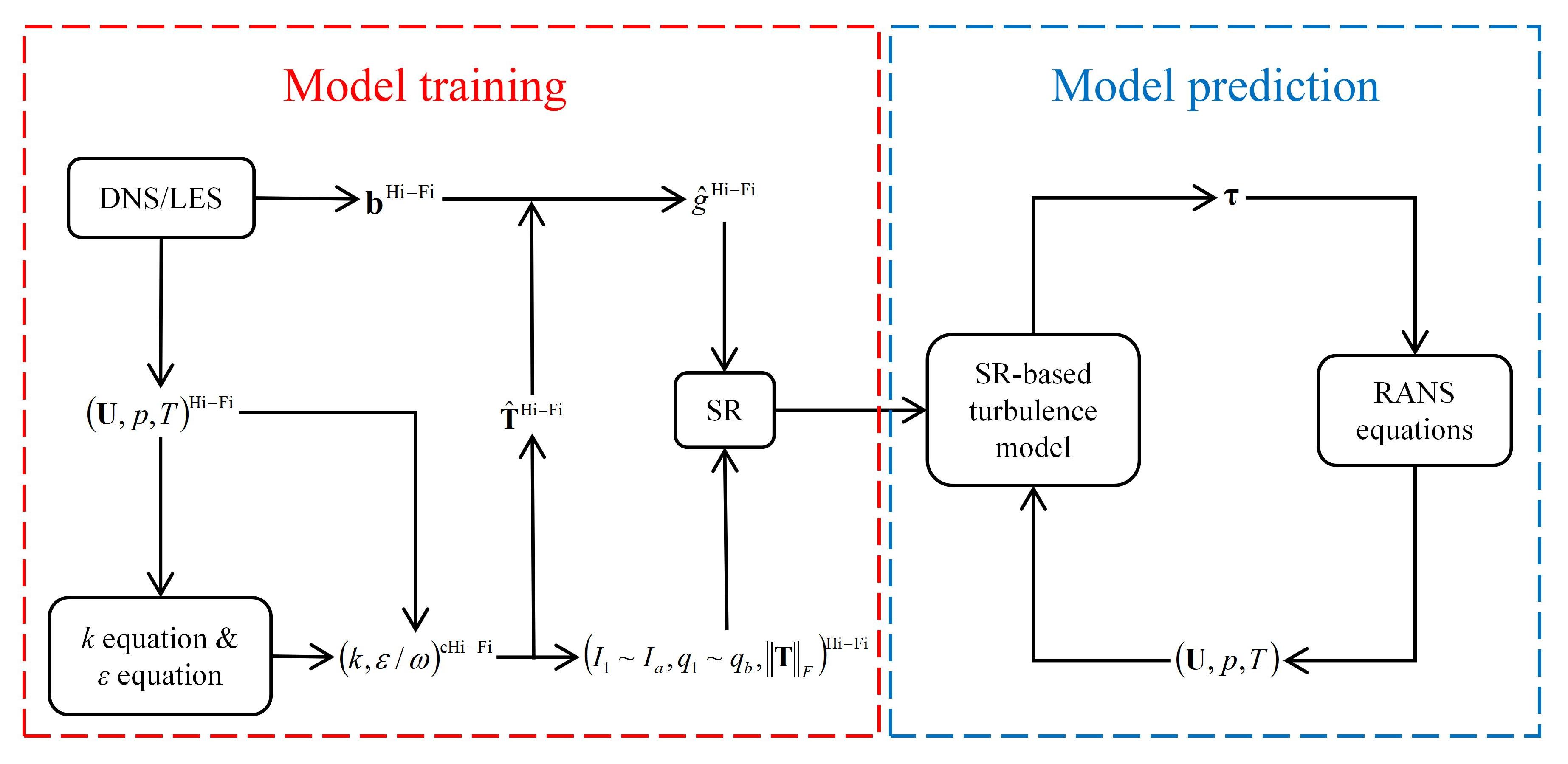} \caption{The framework of our symbolic regression-based turbulence model.}\label{Framework}
\end{figure*}

The methodological framework employed in this investigation is elucidated in detail below. Fig. \ref{Framework} provides a comprehensive schematic representation of our proposed approach.

1. Interpolate high-fidelity DNS or LES data onto the RANS mesh to obtain high-fidelity mean flow quantities $(\mathbf{U}, p, T)^{\mathrm{Hi-Fi}}$ and Reynolds stress field $\mathbf{\tau}^{\mathrm{Hi-Fi}}$.
    
2. Under frozen mean flow conditions, substitute high-fidelity mean flow data $(\mathbf{U}, p, T)^{\mathrm{Hi-Fi}}$ into turbulence model scale transport equations ($k$-equation and $\varepsilon$-equation) to generate compatible turbulence fields $(k, \varepsilon / \omega)^{\mathrm{cHi-Fi}}$.
    
3. Construct input features $\left(I_1 \sim I_{\mathrm{a}}, q_1 \sim q_{\mathrm{b}}, ||\mathbf{T}||_F\right)^{\mathrm{Hi-Fi}}$ and normalized tensor basis $\hat{\mathbf{T}}^{\mathrm{Hi-Fi}}$ based on $(\mathbf{U}, p, T)^{\mathrm{Hi-Fi}}$ and $(k, \varepsilon / \omega)^{\mathrm{cHi-Fi}}$. Simultaneously, calculate the normalized tensor basis coefficients $\hat{g}^{\mathrm{Hi-Fi}}$ according to Eq. (\ref{eq: g_normalized}), using $\mathbf{\tau}^{\mathrm{Hi-Fi}}$ and the normalized tensor bases $\hat{\mathbf{T}}^{\mathrm{Hi-Fi}}$.
    
4. Train machine learning algorithms using $\left(I_1 \sim I_{\mathrm{a}}, q_1 \sim q_{\mathrm{b}}, ||\mathbf{T}||_F\right)^{\mathrm{Hi-Fi}}$ as input and $\hat{g}^{\mathrm{Hi-Fi}}_i$ as output.
    
5. Embed the trained machine learning model into the TCAE code. During the numerical solution iteration process, calculate the Reynolds stress tensor $\mathbf{\tau}$ using mean flow variables $(\mathbf{U}, p, T)$, and incorporate it into the RANS equation system for a coupled solution until the flow field satisfies convergence criteria.

\subsection{Dataset}
\label{sec:Dataset}

\begin{figure*}[htbp]
\centering \includegraphics[width=0.8\textwidth]{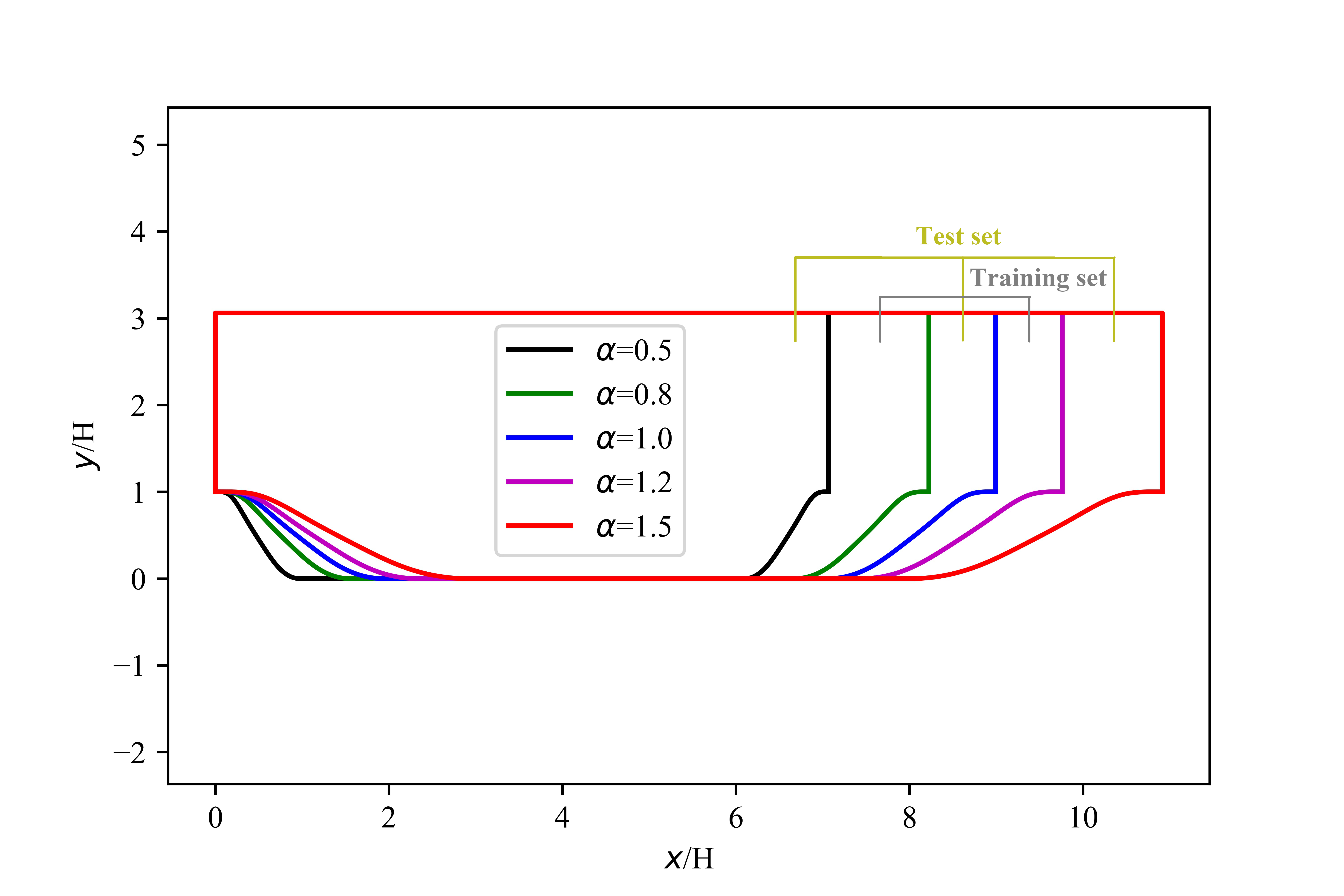} \caption{Parameterized period hill geometries with different $\alpha$.}\label{Parameterized period hill geometries with different}
\end{figure*}

In this study, we employ DNS data of parameterized periodic hill flows from Xiao et al. \cite{xiao_flows_2019} as our training dataset. The geometries of these parameterized periodic hills with varying $\alpha$ values are illustrated in Fig. \ref{Parameterized period hill geometries with different}. For our analysis, we utilize data with $\alpha = 0.8$ and $\alpha = 1.2$ as the training set while reserving $\alpha = 0.5$, $1.0$, and $1.5$ for part of the test set. This experimental design allows us to evaluate our model's performance on interpolated ($\alpha = 1.0$) and extrapolated ($\alpha = 0.5$ and $1.5$) flow conditions.

Table \ref{tab:test_cases} presents the flow scenarios for the whole test case. These test cases enable a comprehensive assessment of the generalizability of our symbolic regression-based turbulence model across multiple levels.

\begin{table*}[htbp]
\caption{\label{tab:test_cases}Test case flow scenarios}
\begin{ruledtabular}
\begin{tabular}{p{3.5cm}p{5cm}p{8cm}}
Test case flow scenario & Flow Characteristics & Description \\ \hline
Periodic hill & 2D, Incompressible, Low Reynolds number, Flow separation induced by bluff body & Separated flow over a bluff body with geometry analogous to the training set. This case evaluates the predictive capability of the symbolic regression model for geometries closely resembling those in the training dataset and can be utilized to assess level 1 generalizability. \\ 
Zero pressure gradient flat plate & 2D, Incompressible, Low Reynolds number & Evaluate whether the symbolic regression turbulence model adversely affects the baseline model's prediction accuracy for wall-bounded flow quantities. This case can be utilized to assess level 2 generalizability. \\
NACA0012 3D airfoil incompressible flow & 3D, Incompressible, High Reynolds number, Flow separation induced by bluff body & Separated flow over a bluff body featuring a geometry and Reynolds number distinctly differs from those in the training set. This case can be employed to assess level 3 generalizability. \\
T106 3D cascade compressible flow and NASA Rotor 37 transonic axial compressor rotor flow & 3D, Compressible, no separation in the T106 case due to the favorable pressure gradient, whereas the NASA Rotor 37 demonstrates separated flow phenomena resulting from the combined effects of bluff-body geometry and adverse pressure gradient conditions & Complex engineering flows significantly differ from those in the training set, representing realistic turbomachinery operating conditions. This case assesses the model's level 4 generalizability. \\
\end{tabular}
\end{ruledtabular}
\end{table*}

\section{Results}\label{sec:Results}

The performance of these turbulence models, as represented by Eq. (\ref{eq: g_hat}), Eq. (\ref{eq: k-epsilon}), Eq. (\ref{eq: gihat_SR}), and Eq. (\ref{eq: KFP}), is thoroughly examined in the subsequent sections.

\subsection{\label{sec:Periodic hill flows}Periodic hill flows}

The computational domain of periodic hill flows is illustrated in Fig. \ref{Parameterized period hill geometries with different}. The simulations are conducted at Reynolds number $Re = 5600$. The detailed mesh configuration employed in the RANS simulations and DNS results can be found in Xiao et al. \cite{xiao_flows_2019}.

\begin{figure*}[htbp]
\centering
\includegraphics[width=1.0\textwidth]{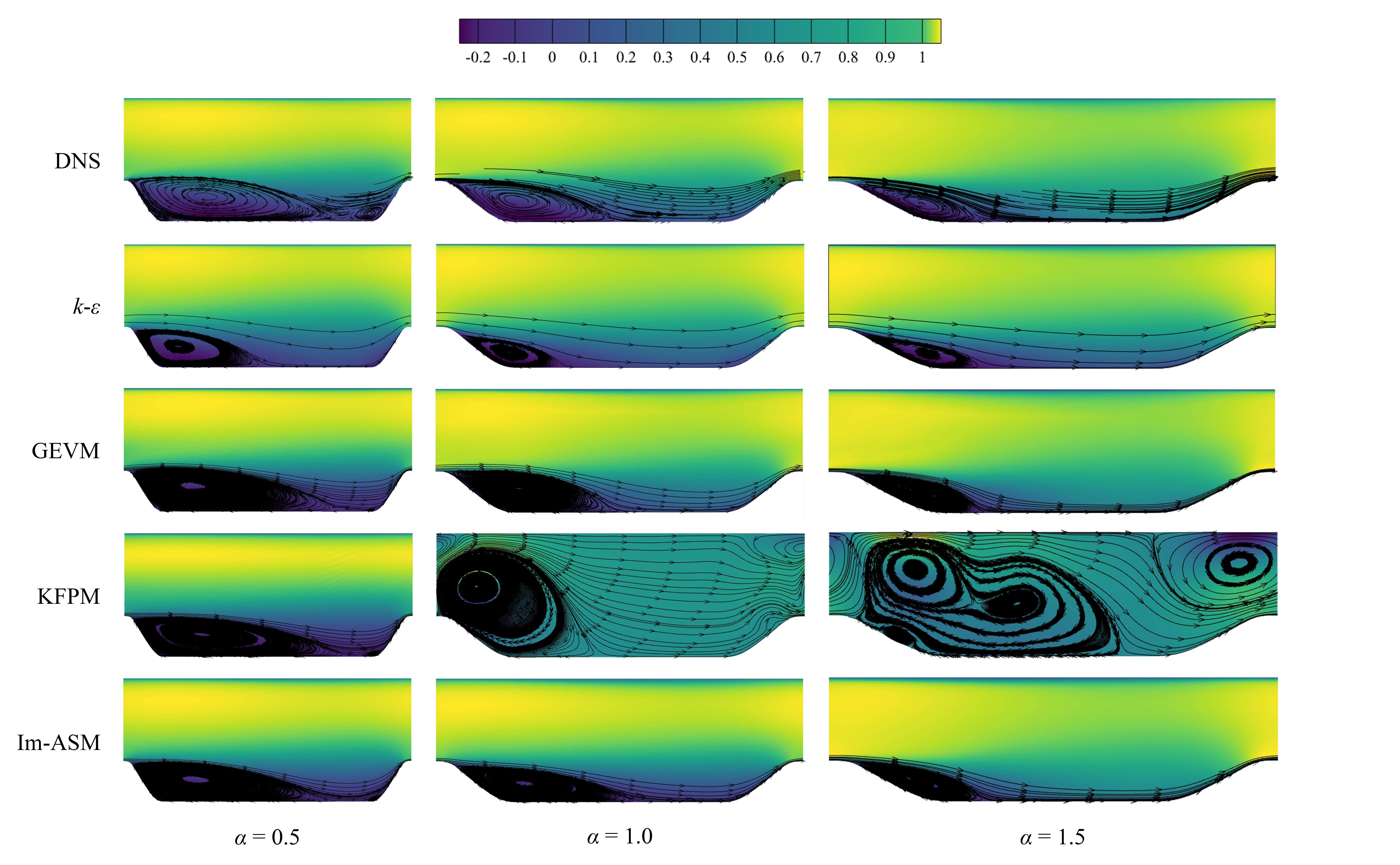}
\caption{\label{Un_SR}Streamlines from various simulation results of the test sets. The contours represent $U_x/U_b$.}
\end{figure*}

The validation dataset for periodic hill flows has been introduced in Section \ref{sec:Dataset}, consisting of cases with $\alpha = 0.5, 1.0, 1.5$ that serve as the validation set for evaluating the performance of our model on periodic hill flow configurations. Fig. \ref{Un_SR} illustrates streamlines derived from various simulation results across the test sets, with contours representing $U_x/U_b$. For the $\alpha = 0.5$ case, the baseline $k$-$\varepsilon$ model significantly underpredicts the separation region compared to DNS results. Although GEVM, KFPM, and Implicit-ASM successfully capture the primary separation vortex in the left portion of the computational domain, all three models substantially underestimate the secondary separation vortex in the right region. In the $\alpha = 1.0$ case, the baseline $k$-$\varepsilon$ model again underpredicts the separation region relative to DNS results. Notably, KFPM produces divergent results, indicating inferior robustness compared not only to GEVM and Implicit-ASM but also to the baseline $k$-$\varepsilon$ model. GEVM demonstrates remarkable accuracy in reproducing vortex morphology comparable to DNS results, while Implicit-ASM predicts a somewhat flattened structure. For the $\alpha = 1.5$ case, the baseline $k$-$\varepsilon$ model, GEVM, and Implicit-ASM all generate vortex structures that closely approximate DNS results. Consistent with the $\alpha = 1.0$ scenario, KFPM again yields divergent outcomes.

\begin{figure}[htbp]
\centering 
\begin{minipage}{0.48\textwidth}
  \centering
  \includegraphics[width=\textwidth]{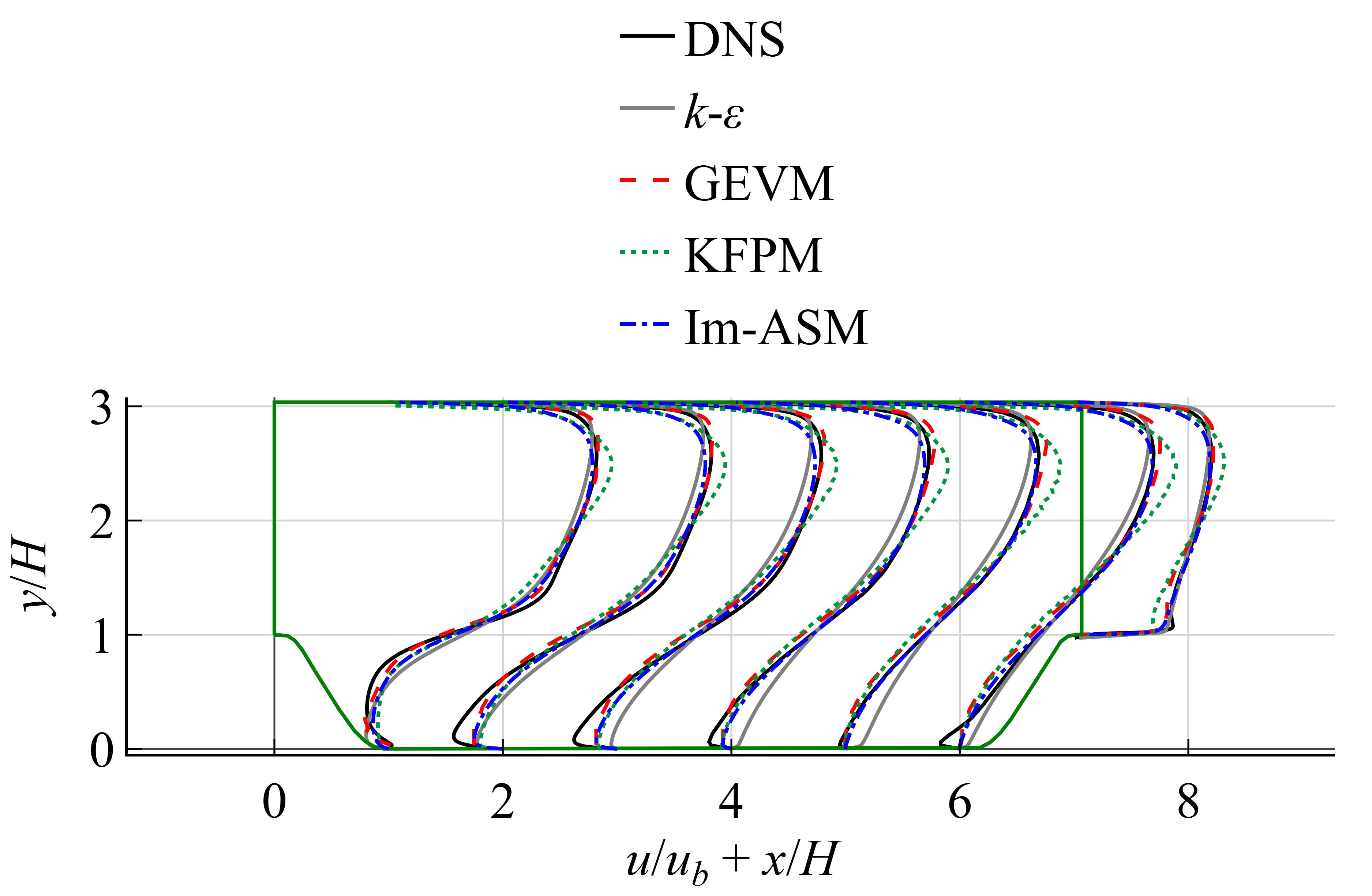}
  \centerline{(a)}
\end{minipage}
\hfill
\begin{minipage}{0.48\textwidth}
  \centering
  \includegraphics[width=\textwidth]{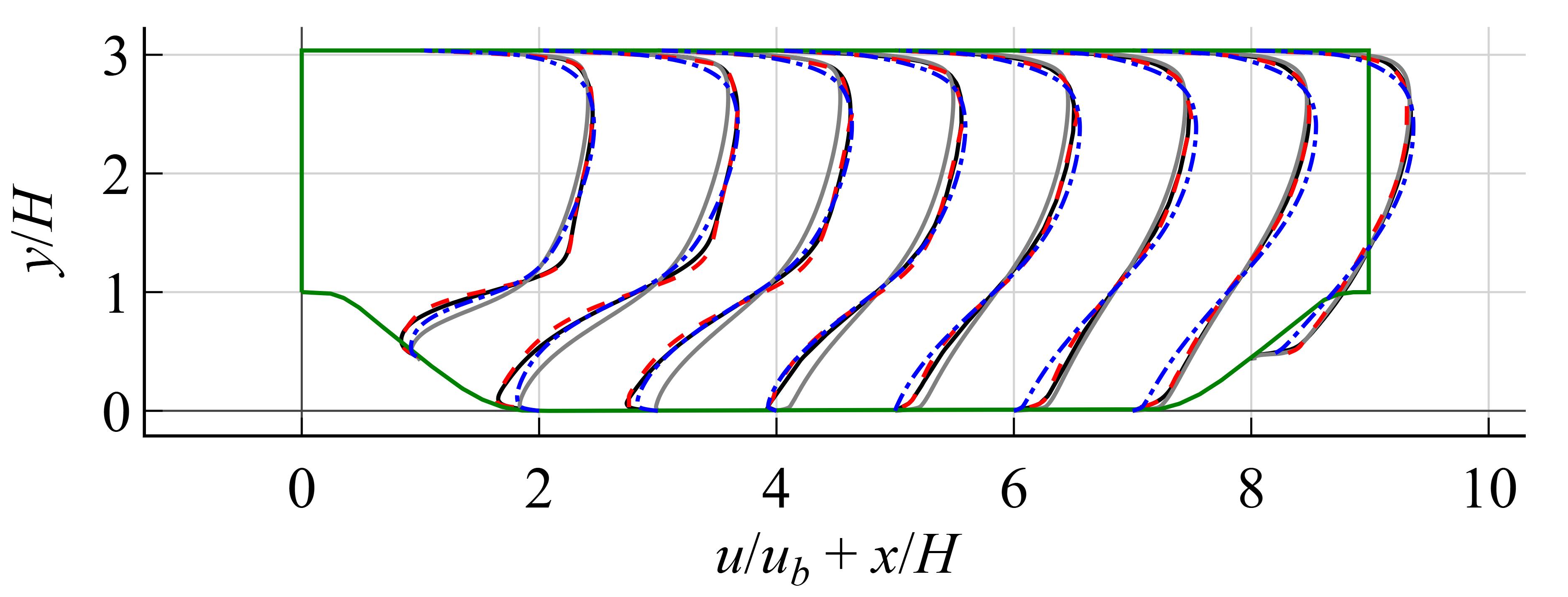}
  \centerline{(b)}
\end{minipage}
\begin{minipage}{0.48\textwidth}
  \centering
  \includegraphics[width=\textwidth]{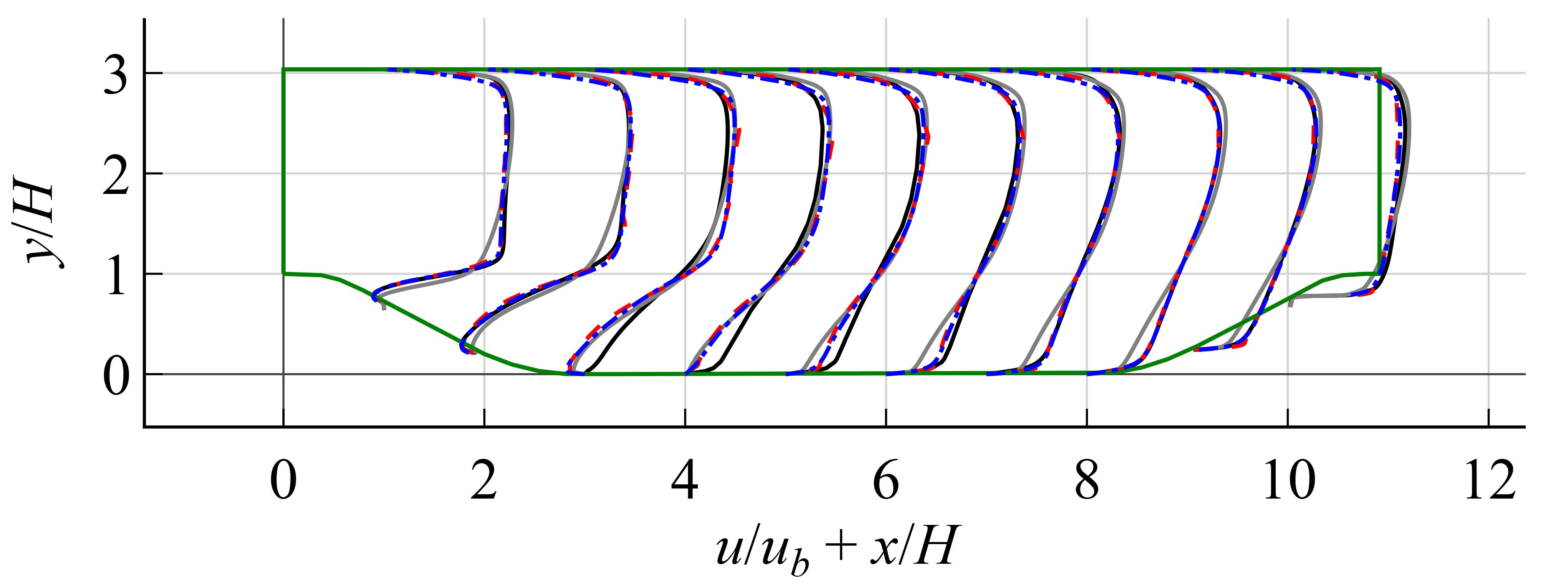}
  \centerline{(c)}
\end{minipage}
\caption{Velocity profiles across the test set from different simulation results. Panels (a), (b), and (c) present the results for the $\alpha = 0.5$, $\alpha = 1.0$, and $\alpha = 1.5$ cases, respectively.}
\label{velocity_profile}
\end{figure}

Fig. \ref{velocity_profile} illustrates the velocity profiles across the test set derived from various simulation results. Panels (a), (b), and (c) correspond to cases where $\alpha = 0.5$, $\alpha = 1.0$, and $\alpha = 1.5$, respectively. Due to convergence issues, KFPM results are presented exclusively for the $\alpha = 0.5$ case, as this method exhibits divergence in both the $\alpha = 1.0$ and $\alpha = 1.5$ scenarios. For the $\alpha = 0.5$ case, both Implicit-ASM and KFPM demonstrate slight deviations from DNS data in the vicinity of the top wall region. Specifically, KFPM predicts elevated velocity magnitudes near $y/H = 2.5$ compared to DNS results. In the region proximal to the bottom wall, GEVM, KFPM, and Implicit-ASM generally outperform the baseline $k$-$\varepsilon$ model. Regarding the $\alpha = 1.0$ case, although Implicit-ASM performs marginally less accurately than GEVM, it nevertheless demonstrates superior performance relative to the baseline $k$-$\varepsilon$ model overall. For the $\alpha = 1.5$ case, Implicit-ASM and GEVM exhibit comparable performance, both surpassing the baseline $k$-$\varepsilon$ model in accuracy.

Table \ref{Table: Comparison of MSE across various models in periodic hills flow Test cases.} presents a comprehensive comparison of mean square error (MSE) across various turbulence models in the test cases. For the $\alpha = 0.5$ configuration, GEVM, KFPM, and Implicit-ASM all demonstrate superior predictive accuracy compared to the baseline $k$-$\varepsilon$ model, although KFPM exhibits only marginal improvement over the baseline. In the $\alpha = 1.0$ and $\alpha = 1.5$ configurations, both GEVM and KFPM outperform the baseline $k$-$\varepsilon$ model, with GEVM demonstrating significantly enhanced performance relative to Implicit-ASM.

\begin{table}[htbp]
\caption{\label{Table: Comparison of MSE across various models in periodic hills flow Test cases.}Comparison of MSE across various models in test cases.}
\begin{ruledtabular}
\begin{tabular}{ccccc}
 & $k$-$\epsilon$ & GEVM & KFPM & Im-ASM \\ \hline
$\alpha=0.5$ & $6.25\times10^{-3}$ & $2.42\times10^{-3}$ & $5.10\times10^{-3}$ & $2.41\times10^{-3}$ \\
$\alpha=1.0$ & $4.36\times10^{-3}$ & $6.45\times10^{-4}$ & - & $3.09\times10^{-3}$ \\
$\alpha=1.5$ & $2.41\times10^{-3}$ & $7.92\times10^{-4}$ & - & $1.96\times10^{-3}$ \\
\end{tabular}
\end{ruledtabular}
\end{table}

\begin{figure*}[htbp]
\centering
\includegraphics[width=0.8\textwidth]{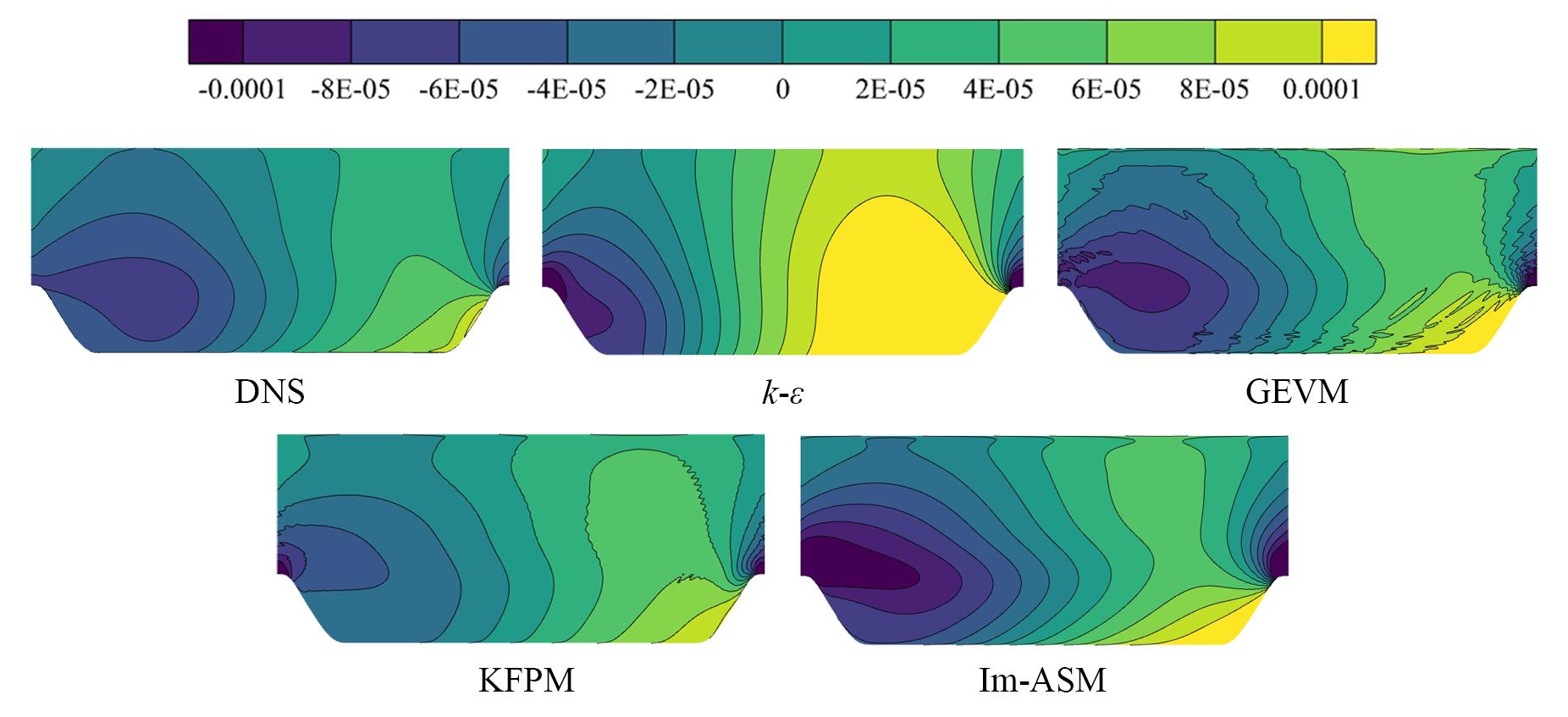}
\caption{\label{PH_P_0.5}Pressure (Pa) distributions obtained from various simulation results of $\alpha = 0.5$ case.}
\end{figure*}

Fig. \ref{PH_P_0.5} illustrates the pressure distributions derived from various simulation results for the case where $\alpha = 0.5$. The results demonstrate that explicit models such as GEVM and KFPM can produce discontinuous pressure distributions, whereas the Implicit-ASM model predicts a smooth pressure profile.

Based on our analysis, the GEVM and Implicit-ASM models demonstrate superior performance compared to the baseline $k$-$\varepsilon$ model in this section. However, it is noteworthy that GEVM exhibits considerable non-unsmoothness in pressure field predictions. While the KFPM model shows marginally better performance than the baseline $k$-$\varepsilon$ model under the $\alpha = 0.5$ condition, its robustness is questionable given the simulation results at $\alpha = 1.0$ and $\alpha = 1.5$ conditions. Due to convergence difficulties, KFPM results will be omitted from subsequent sections. Therefore, we conclude that the Implicit-ASM model successfully achieves generalization level 1.

\subsection{\label{sec:Zero pressure gradient flat plate flow}Zero pressure gradient flat plate flow}

\begin{figure*}[htbp]
\centering \includegraphics[width=0.5\textwidth]{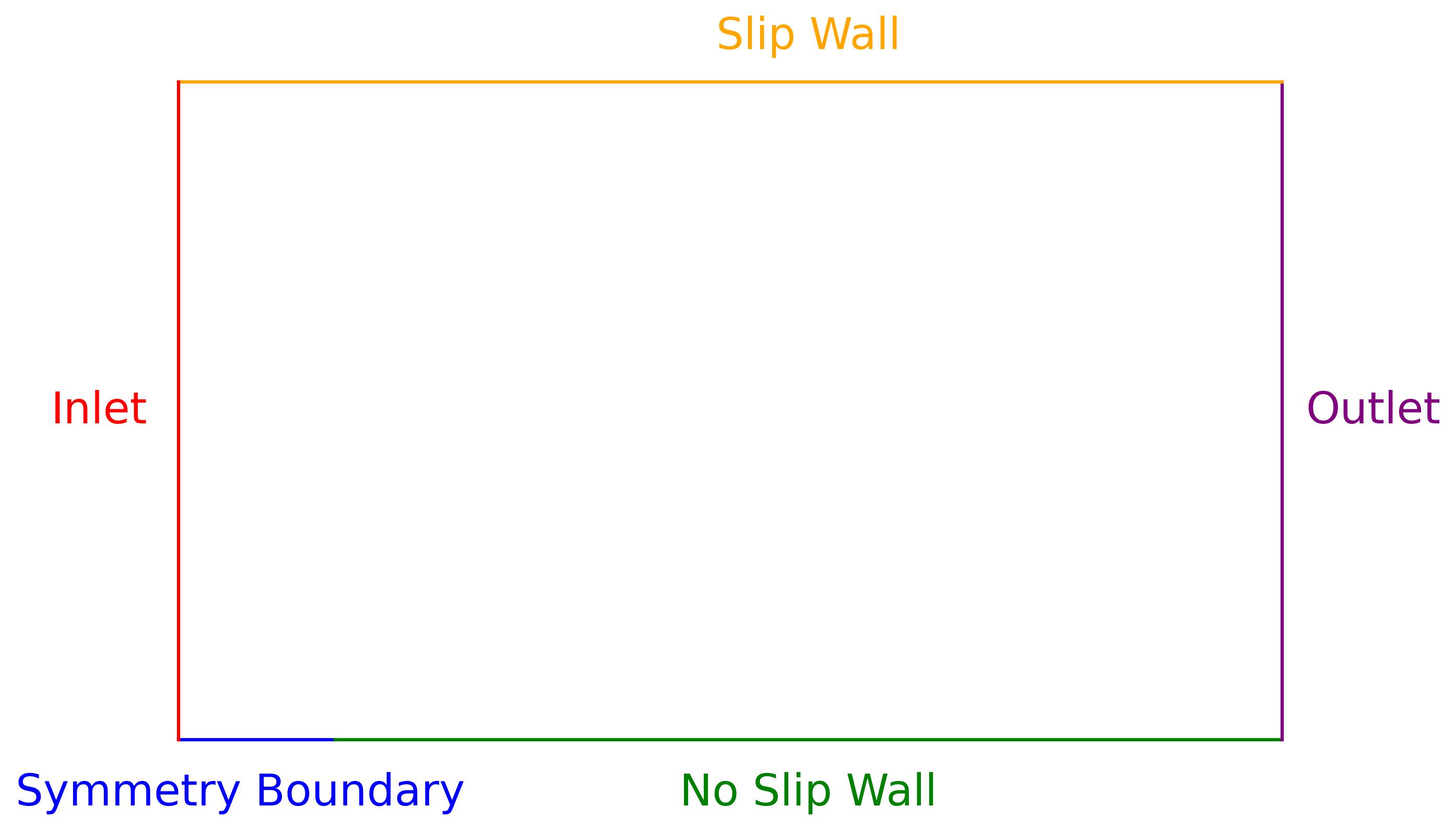} \caption{Computational domain for the zero pressure gradient flat plate flow simulation.}\label{flat_plate}
\end{figure*}

The zero pressure gradient flat plate flow is characterized by a Reynolds number, based on the plate length, of $Re = 5 \times 10^6$. The computational domain is illustrated in Fig. \ref{flat_plate}.

\begin{figure}[htbp]
\centering \includegraphics[width=0.48\textwidth]{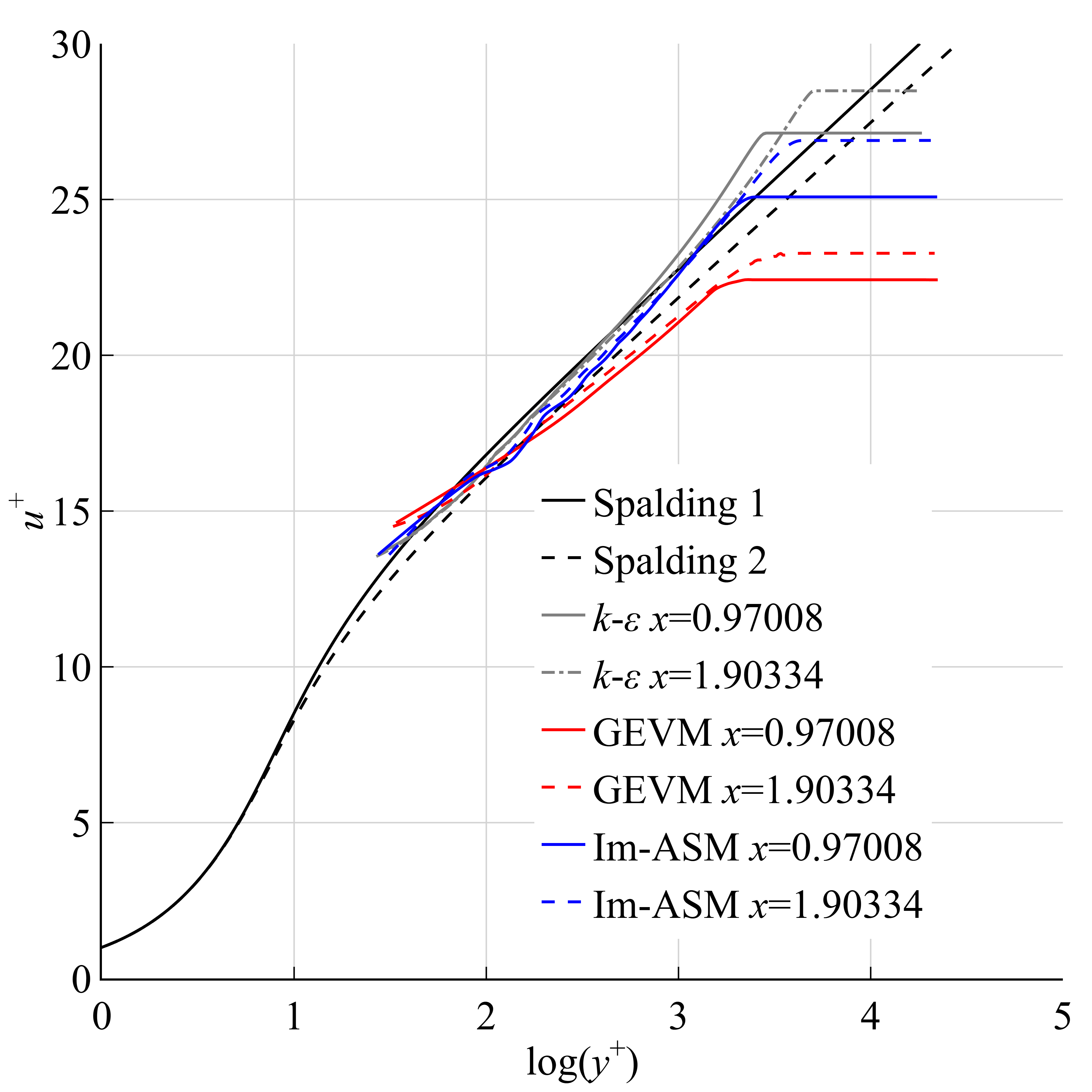} \caption{Comparative analysis of law of the wall results.}\label{law_of_the_wall}
\end{figure}

Fig. \ref{law_of_the_wall} presents a comparative analysis between the computational results obtained from the baseline $k$-$\epsilon$ model, GEVM, and Implicit-ASM, juxtaposed with the canonical law of the wall formulation proposed by Spalding \cite{spalding_single_1961}. Spalding 1 and Spalding 2 represent two distinct results with different coefficient selections: Spalding 1 employs a von Kármán coefficient of 0.41 and an integration coefficient of 5.5 while Spalding 2 utilizes a von Kármán coefficient of 0.40 and an integration coefficient of 5.0. The analysis demonstrates that the baseline $k$-$\epsilon$ model yields predictions more closely aligned with Spalding 1 data, albeit with some discrepancies in the high $y^+$ regime. In contrast, GEVM predictions show stronger correspondence with Spalding 2 data. The Implicit-ASM produces results that generally fall between Spalding 1 and Spalding 2 data across most regimes, though exhibiting some un-smooth in its curve profiles. Overall, both GEVM and Implicit-ASM deliver more reasonable results compared to the baseline $k$-$\epsilon$ model, successfully achieving generalization level 2.

Based on the results presented in this section, GEVM demonstrates slight deviations from the Spalding 2 data. While the Implicit-ASM results exhibit minor fluctuations within a limited regime, they generally fall between the Spalding 1 and Spalding 2 datasets. Consequently, we conclude that Implicit-ASM achieves generalization level 2.

\subsection{\label{sec:Three-dimensional incompressible flow around a NACA0012 airfoil}Three-dimensional incompressible flow around a NACA0012 airfoil.}

In this section, we evaluate the performance of our symbolic regression-based turbulence models on the three-dimensional incompressible flow around a NACA0012 airfoil. The simulation is conducted at $Re = 3 \times 10^6$. The computation mesh configuration can be found in our previous study \cite{ji2025enhancinggeneralizabilitymachinelearning}. The spanwise dimension extends to 0.2 chord lengths and is discretized with 10 grid points. Experimental data for validation are obtained from Ladson et al. \cite{ladson1987pressure}.

\begin{figure*}[htbp]
\centering 
\begin{minipage}{0.3\textwidth}
  \centering
  \includegraphics[width=\textwidth]{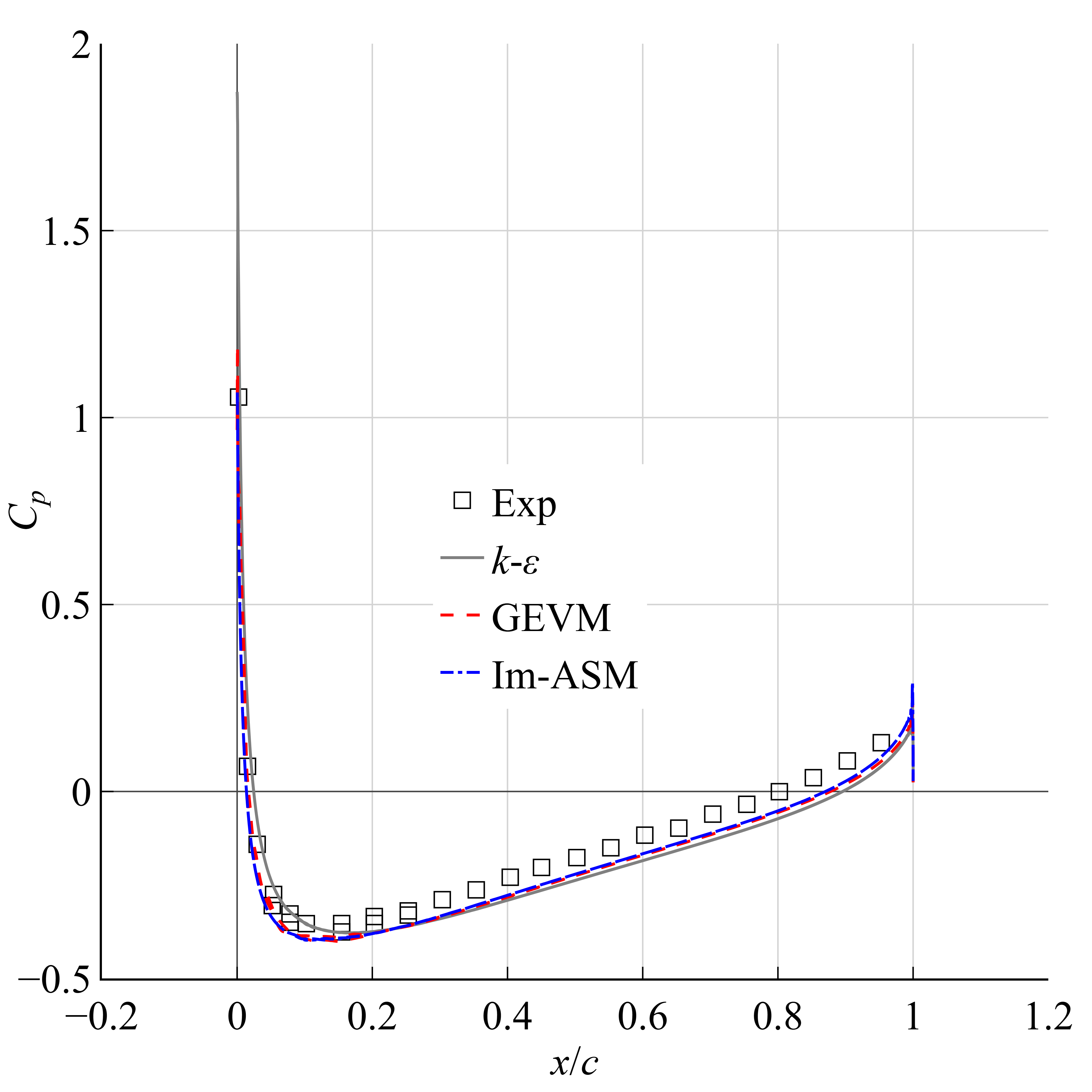}
  \centerline{(a)}
\end{minipage}
\begin{minipage}{0.3\textwidth}
  \centering
  \includegraphics[width=\textwidth]{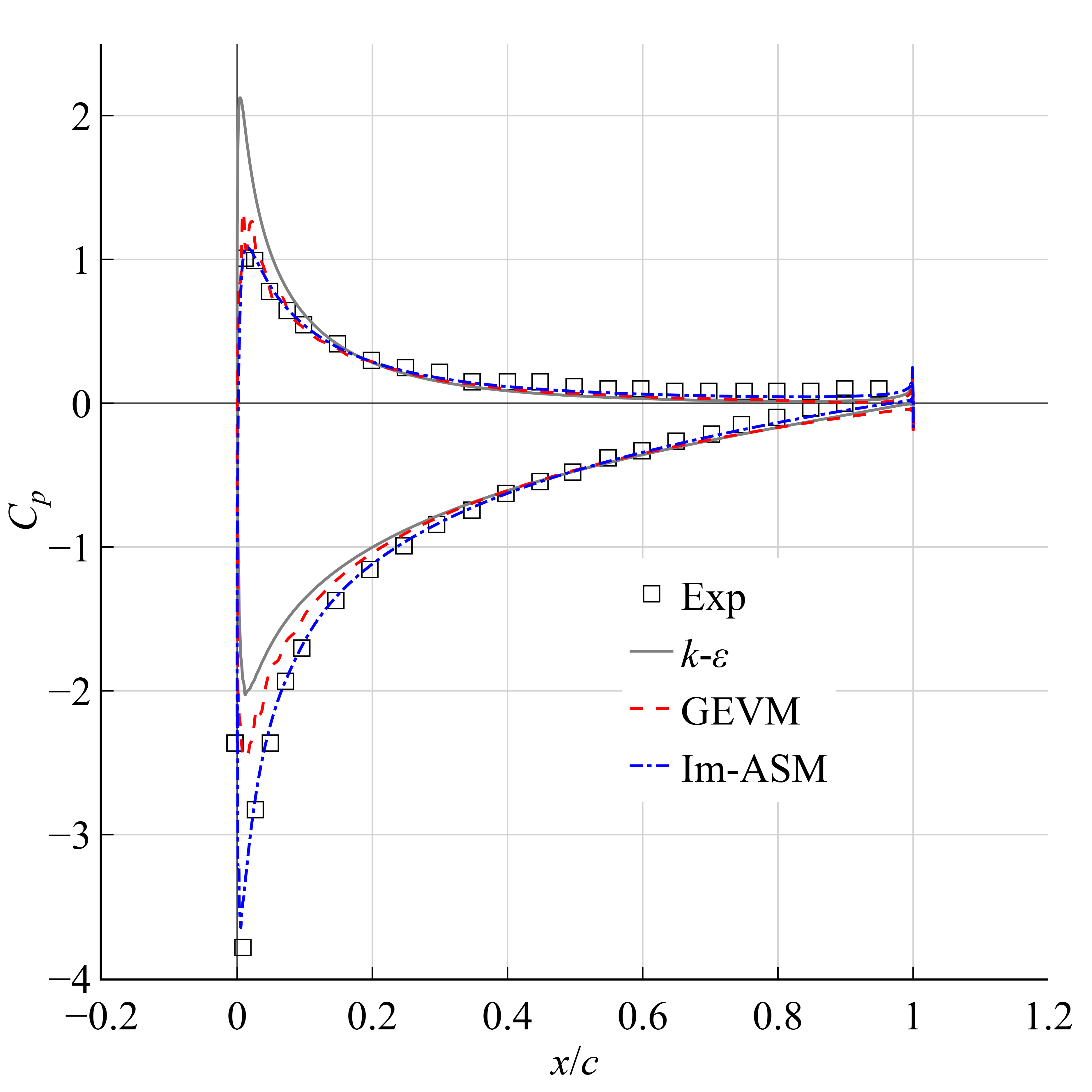}
  \centerline{(b)}
\end{minipage}
\begin{minipage}{0.3\textwidth}
  \centering
  \includegraphics[width=\textwidth]{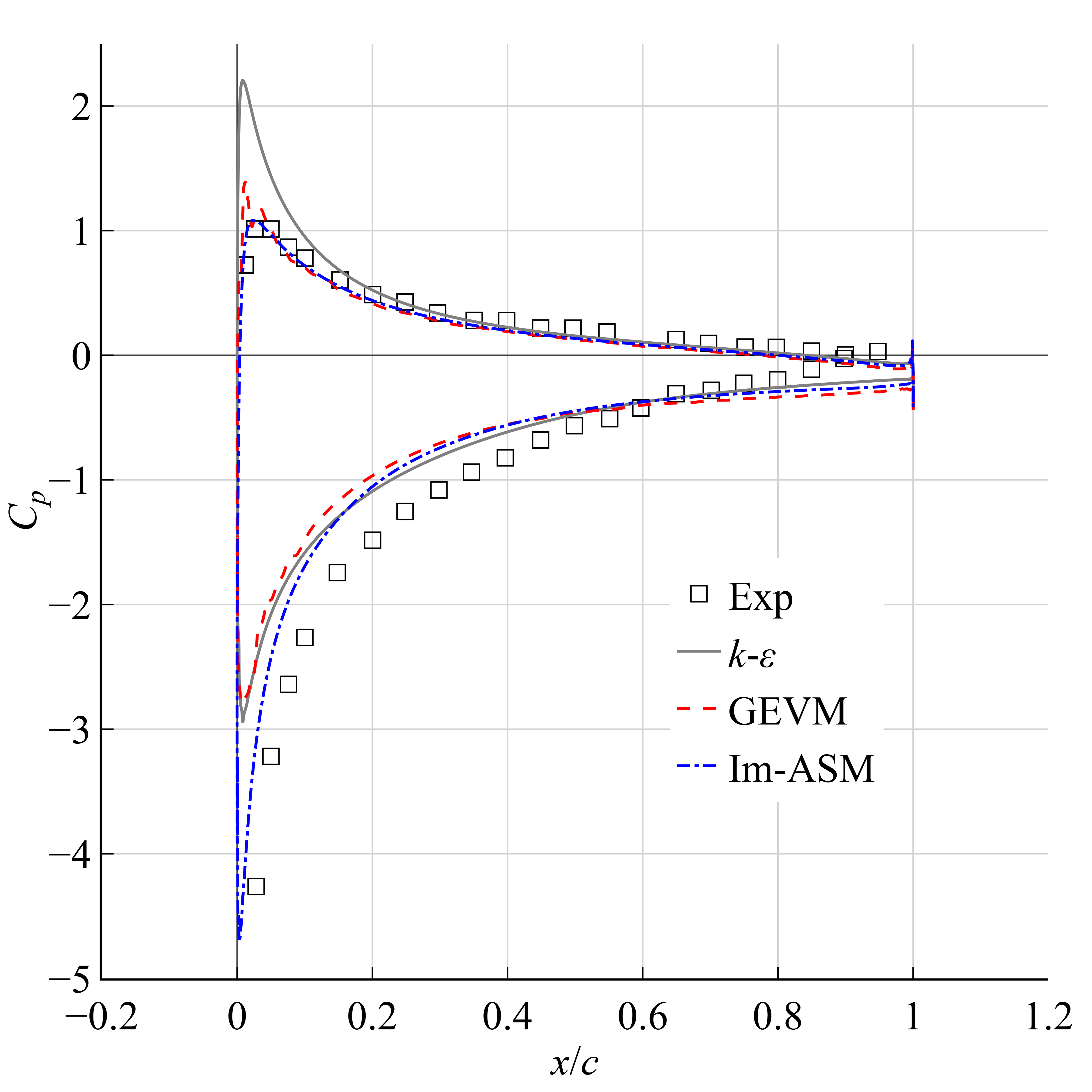}
  \centerline{(c)}
\end{minipage}
\caption{A comparative analysis of pressure coefficient distributions along the NACA0012 airfoil surface at various angles of attack. Panels (a), (b), and (c) illustrate the pressure distributions at incidence angles of $0^{\circ}$, $10^{\circ}$, and $15^{\circ}$, respectively.}
\label{NACA0012}
\end{figure*}

Fig. \ref{NACA0012} presents a comparative analysis of pressure coefficient distributions along the NACA0012 airfoil surface at various angles of attack. Panels (a), (b), and (c) illustrate the pressure distributions at incidence angles of $0^{\circ}$, $10^{\circ}$, and $15^{\circ}$, respectively. At $0^{\circ}$ angle of attack, both GEVM and Im-ASM demonstrate improved prediction accuracy at the leading edge ($x/c=0$), with Im-ASM closely approximating the experimental results at this location. At $10^{\circ}$ angle of attack, Im-ASM yields highly satisfactory results when compared to experimental data. Although the GEVM model outperforms the baseline $k$-$\epsilon$ model, it exhibits some unsmoothness in the leading edge region of the airfoil. Similarly, at $15^{\circ}$ angle of attack, Im-ASM performs exceptionally well in the leading edge region, while GEVM continues to display unsmoothness behavior in this area.

Based on the results presented in this section, GEVM exhibits non-smooth predictions in certain regions, consistent with observations from previous sections. The Implicit-ASM demonstrates significantly superior predictive performance compared to both the baseline $k$-$\varepsilon$ model and GEVM. Consequently, we conclude that Implicit-ASM achieves generalization level 3.

\subsection{Three-dimensional compressible turbine cascade flow}
\label{sec:Three-dimensional compressible turbine cascade flow}

The T106 is a representative low-pressure turbine configuration in which 30-40\% of aerodynamic losses are attributed to interactions between the blade profile and the end-wall boundary layer. This interaction generates complex flow phenomena near the T106 end-wall region, effectively simulating the intricate flow characteristics observed in operational turbine environments. Turbine flows generally exhibit favorable pressure gradients, which conventional turbulence models can predict with greater accuracy compared to the adverse pressure gradients typical of compressor flows. Consequently, this case provides an ideal benchmark for evaluating turbulence model performance under conditions of favorable pressure gradient without flow separation. In this investigation, we simulate the three-dimensional T106 cascade flow at an outlet Reynolds number of $Re = 120,000$ and Mach number of $Ma = 0.59$. The experimental data for validation are sourced from Duden and Fottner \cite{duden1997influence}.

\begin{figure}[htbp]
\centering 
\includegraphics[width=0.48\textwidth]{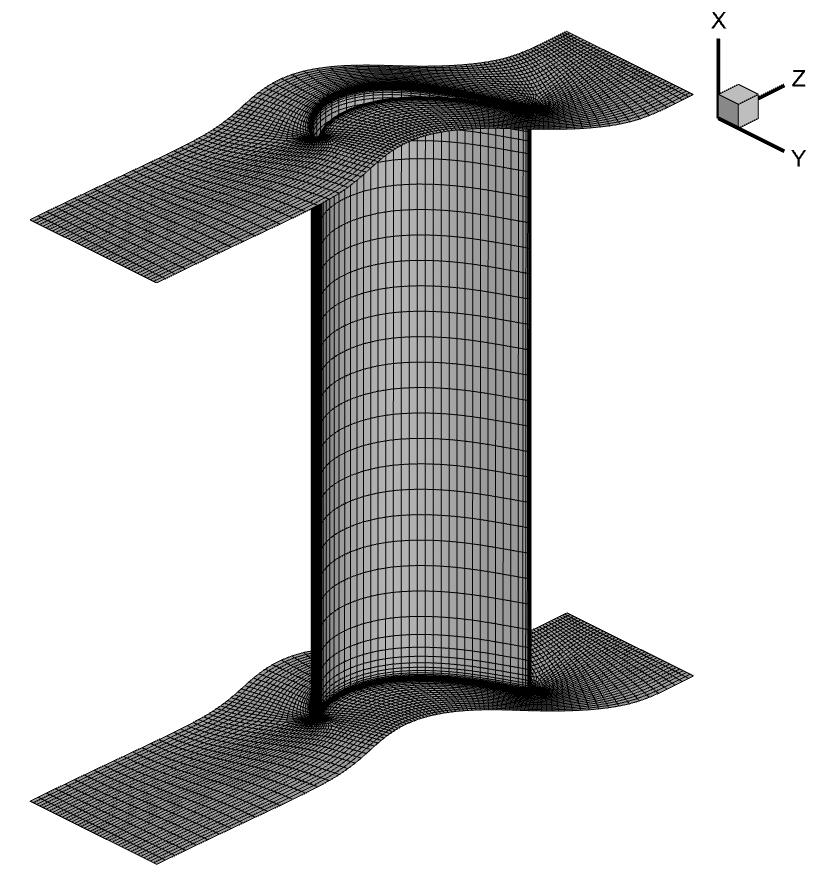} 
\caption{The computational domain and mesh utilized for the T106 cascade simulation.}
\label{T106_mesh}
\end{figure}

Fig. (\ref{T106_mesh}) illustrates the computational domain and mesh configuration employed for the T106 turbine cascade simulation. The predominant flow direction proceeds from the negative to positive orientation along the z-axis.

\begin{figure}[htbp]
\centering 
\includegraphics[width=0.48\textwidth]{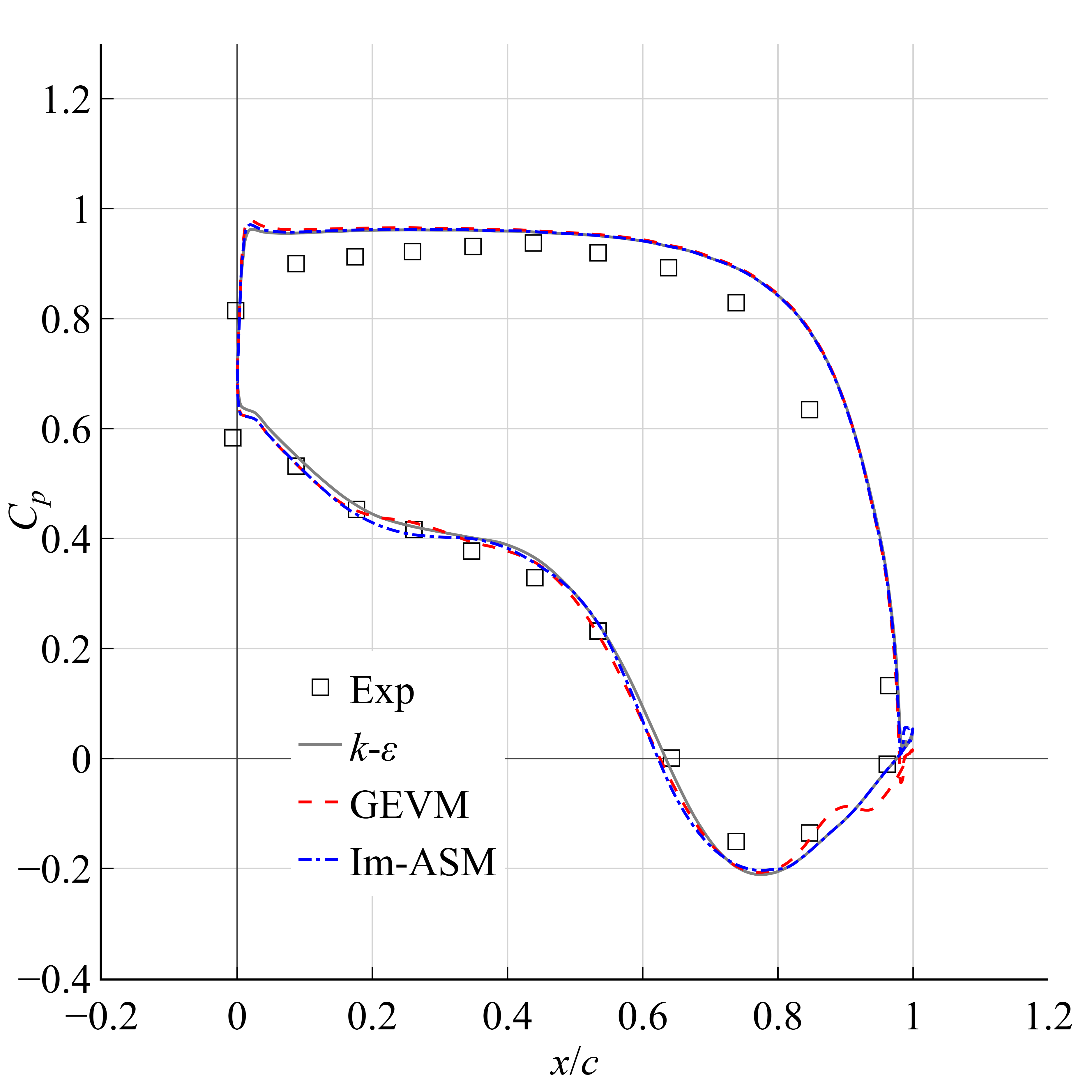} 
\caption{The pressure coefficient distribution along the end-wall surface of the T106 turbine blade.}
\label{T106_Cp}
\end{figure}

Fig. (\ref{T106_Cp}) illustrates the pressure coefficient distribution along the end-wall surface of the T106 turbine blade. The results demonstrate that all models perform well when compared to the experimental data, except GEVM, which predicts irregular behavior near the trailing edge on the suction side.

\begin{figure*}[htbp]
\centering 
\includegraphics[width=0.6\textwidth]{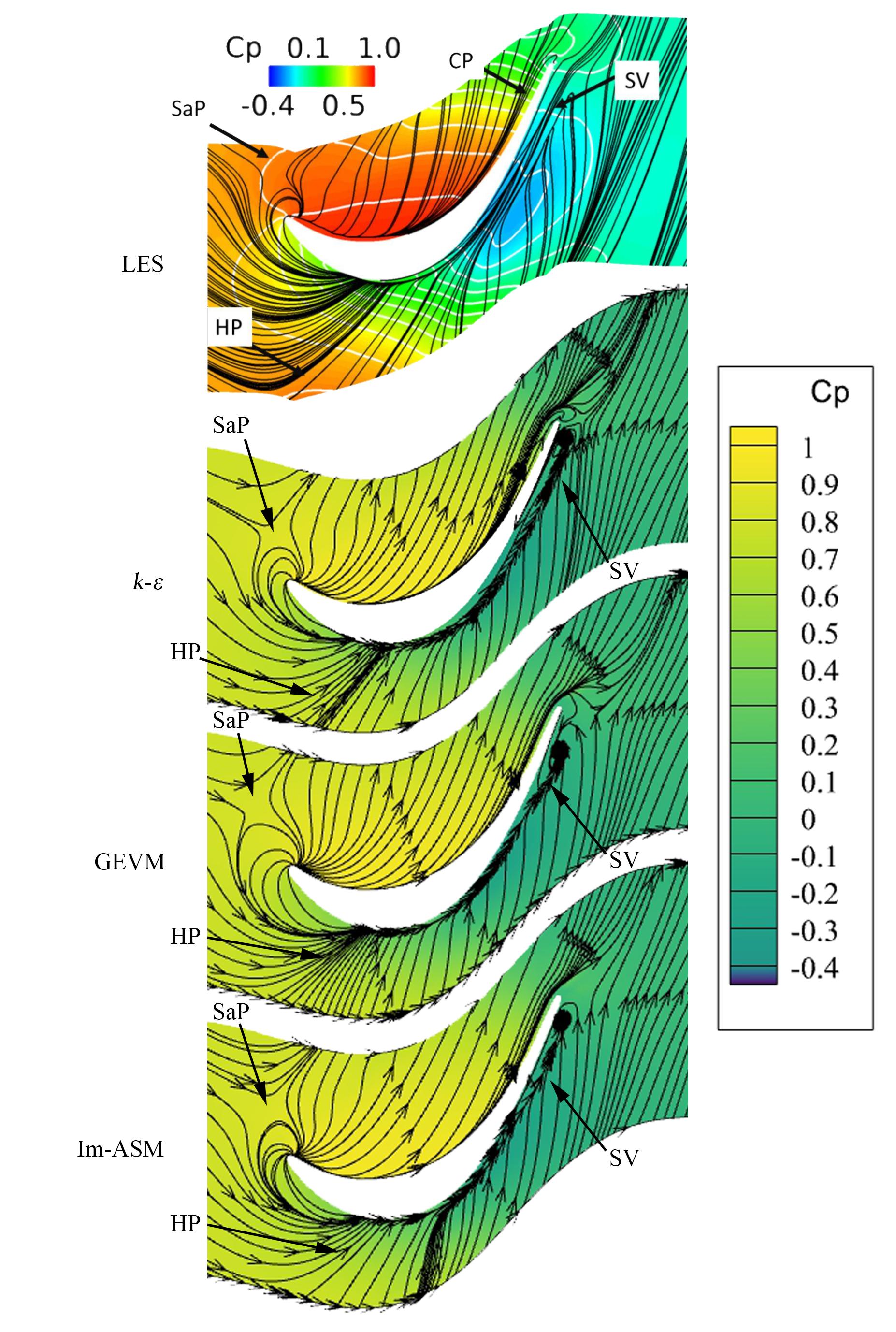} 
\caption{End-wall streamline and flow structure for different results. The LES results are derived from the work of Pichler et al. \cite{pichler_and_2018}. The notation ``Sap'' denotes a saddle point, while ``HP'' represents horseshoe vortex formations. The designation ``SV'' indicates regions characterized by complex interactions among corner, passage, and horseshoe vortices.}
\label{T106_streamline}
\end{figure*}

Fig. (\ref{T106_streamline}) shows the End-wall streamline and flow structure for different results. The LES results are derived from the work of Pichler et al. \cite{pichler_and_2018}. The notation ``Sap'' denotes a saddle point, while ``HP'' represents horseshoe vortex formations. The designation ``SV'' indicates regions characterized by complex interactions among corner, passage, and horseshoe vortices. It is demonstrated that our computational results exhibit strong consistency among themselves and show substantial agreement with the LES results.

\subsection{Transonic axial compressor rotor}
\label{sec:Transonic_axial_compressor_rotor}

\begin{figure}[htbp]
\centering 
\includegraphics[width=0.48\textwidth]{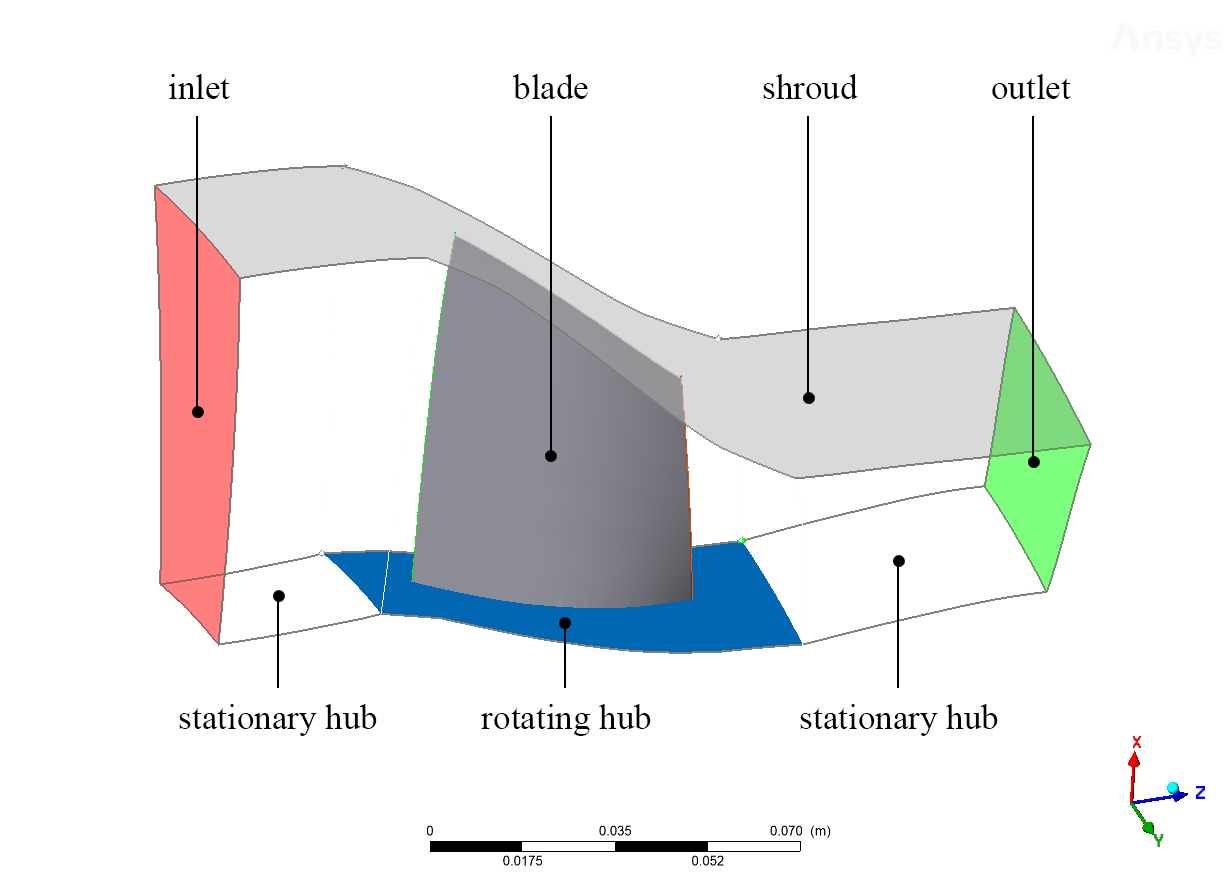} 
\caption{Three-dimensional computational domain of a single blade of NASA Rotor 37.}
\label{Rotor_37}
\end{figure}

In this section, we evaluate the generalizability of our symbolic regression model through an analysis of the NASA Rotor 37, a transonic axial compressor rotor. The design specifications for this rotor are presented in Table \ref{Table: Rotor37}. This compressor rotor employs shock wave mechanisms to enhance fluid pressure and operates at a high rotation speed (17188.7 rpm). The rotor exhibits a high total pressure ratio, classifying it as a sophisticated engineering-level turbulent flow system. Consequently, it provides an ideal test case for assessing the level 4 generalizability of our symbolic regression turbulence model. Fig. \ref{Rotor_37} illustrates the three-dimensional computational domain for a single blade of NASA Rotor 37. Experimental data for this rotor were obtained from Suder \cite{suder_experimental_1996}. Additional computational methodologies and parameters are documented in our previous publication \cite{ji_tensor_2024}.

\begin{table}[htbp]
\caption{\label{Table: Rotor37}Main design parameters for NASA Rotor 37.}
\begin{ruledtabular}
\begin{tabular}{cc}
Design parameters & NASA Rotor 37 \\ \hline
Rotor total pressure ratio & 2.106 \\
Mass flow rate & 20.188 kg/s \\
Rotor wheel speed & 17188.7 rmp \\
Number of rotor blades & 36 \\
\end{tabular}
\end{ruledtabular}
\end{table}

\begin{figure}[htbp]
\centering
\includegraphics[width=0.48\textwidth]{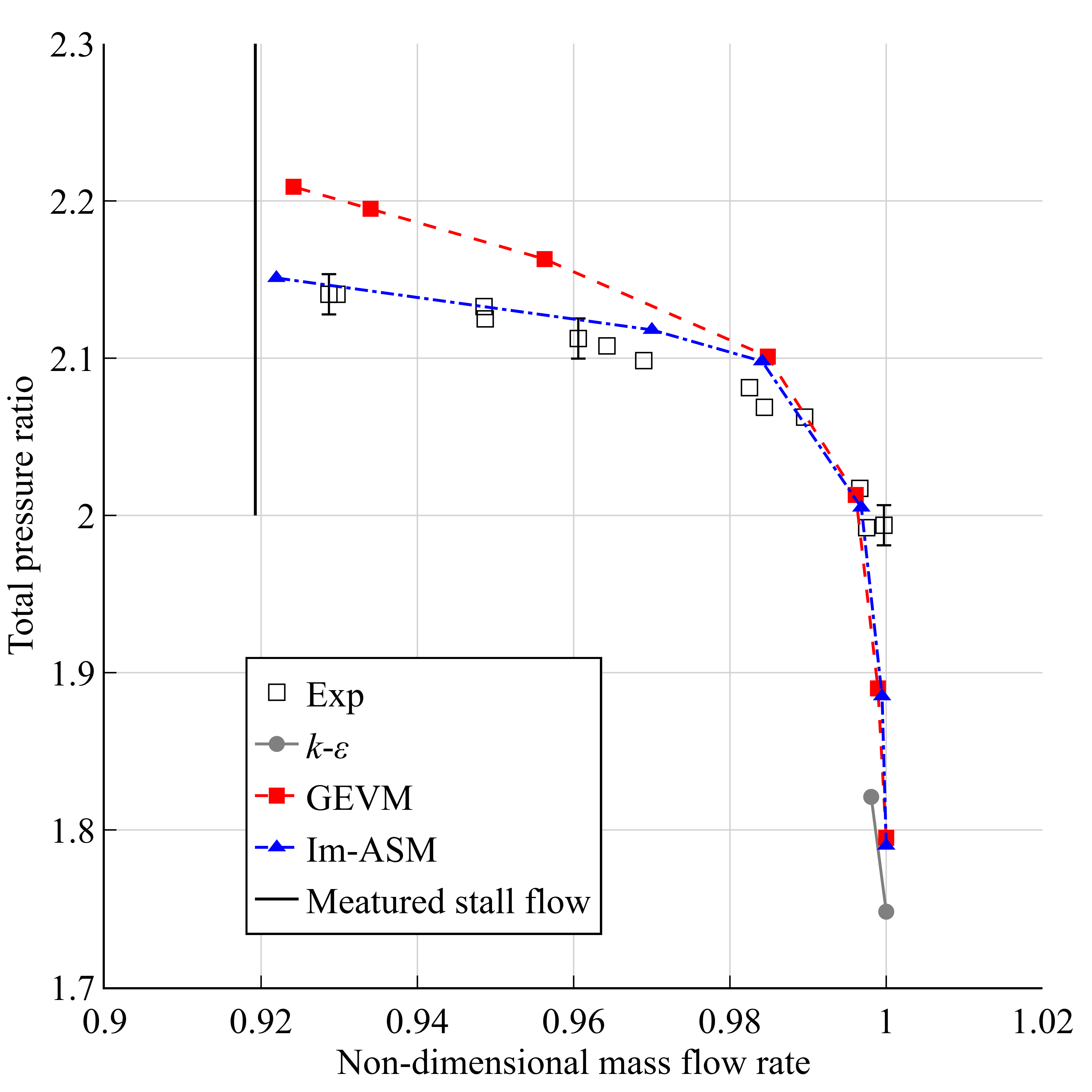}
\caption{Non-dimensional mass flow rate versus total pressure ratio characteristics of NASA Rotor 37.}
\label{fig:rotor37_pressure_ratio}
\end{figure}

Fig. (\ref{fig:rotor37_pressure_ratio}) illustrates the relationship between non-dimensional mass flow rate and total pressure ratio characteristics of NASA Rotor 37. The baseline $k$-$\varepsilon$ model demonstrates limited capability, achieving convergence only in proximity to blockage conditions. The GEVM approach substantially overpredicts the total pressure ratio across most of the operating regime, with results extending well beyond the uncertainty band. In contrast, the Im-ASM method yields significantly more accurate predictions, with the majority of data points falling within the established uncertainty band.

\begin{figure*}[htbp]
\centering
\includegraphics[width=0.7\textwidth]{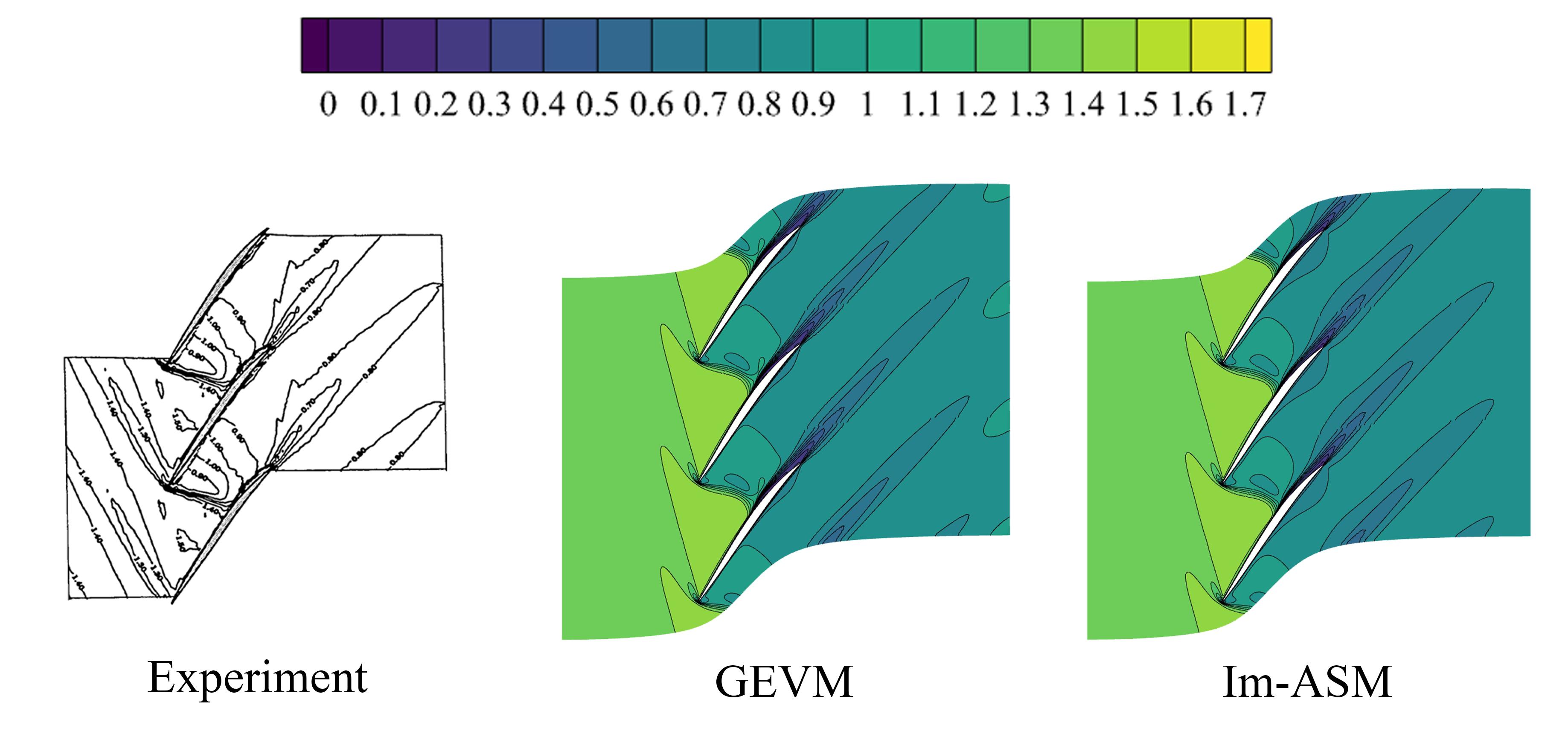}
\caption{Relative Mach number contours on 98\% measured chock flow condition of NASA Rotor 37 at 70\% span.}
\label{Relative_Mach_number}
\end{figure*}

Fig. (\ref{Relative_Mach_number}) illustrates the relative Mach number contours at 98\% measured choked flow condition for NASA Rotor 37 at 70\% span. The results demonstrate that while minor differences in shock wave morphology are evident, the shock location in our numerical simulations closely corresponds to that observed in experimental data. Furthermore, the computational model accurately captures flow separation phenomena and blade wake characteristics when compared to experimental measurements. Notably, the simulated flow field exhibits smooth solution characteristics without numerical artifacts.

Based on the results presented in Sec. \ref{sec:Three-dimensional incompressible flow around a NACA0012 airfoil} and this section, the GEVM exhibits certain un-smoothness in the pressure coefficient distribution for the T106 turbine cascade. In contrast, the Implicit-ASM demonstrates significantly superior performance in predicting the NASA Rotor 37 non-dimensional mass flow rate versus total pressure ratio characteristics compared to both the baseline $k$-$\varepsilon$ model and GEVM. Consequently, we conclude that the Implicit-ASM achieves a generalization level of 4.

\section{\label{sec: Discussion}Discussion}

The key advantage of our Im-ASM approach lies in its consideration of the interaction between the non-dimensional Reynolds stress deviatoric tensor $\mathrm{b}$ and the non-dimensional velocity gradient tensor $\hat{\mathrm{U}}$ as presented in Eq. (\ref{eq:non-dimensional tensor version b}). This interaction represents the production term of $\mathrm{b}$. Given that the non-dimensional Reynolds stress deviatoric tensor $\mathrm{b}$ characterizes the shape of turbulence \cite{lumley_return_1977}, this term effectively captures the contribution of the shape of local turbulence produced by the mean flow field. The unsmoothness observed in the prediction results, illustrated in Fig. (\ref{PH_P_0.5}) for the two explicit turbulence models, can likely be attributed to their failure to incorporate the production term of $\mathrm{b}$.

Although the original formulation of the algebraic stress model proposed by Rodi (Eq. (\ref{eq:ASM})) is derived from precise physical assumptions and rigorous mathematical processes, the numerical robustness of Rodi's model is not necessarily guaranteed by its physical derivation. Similarly, the KFPM presented in Eq. (\ref{eq: KFP}) faces comparable limitations. In contrast, data-driven approaches, such as machine learning, offer promising avenues for developing turbulence models with more robust numerical structures while relaxing the constraints imposed by strict physical assumptions.

\section{\label{sec: Conclusion}Conclusion}

We propose a symbolic regression-based implicit algebraic stress turbulence model by incorporating the production term into the representation of the non-dimensional Reynolds stress deviatoric tensor. Comparative analysis with the baseline $k$-$\varepsilon$ model, general effective-viscosity model, and a model constructed using the kinetic Fokker-Planck equation demonstrates that the Implicit-ASM not only achieves higher accuracy than the baseline $k$-$\varepsilon$ model but also predicts smoother flow fields than both explicit models (GEVM and KFPM). Additionally, the Implicit-ASM exhibits superior robustness compared to KFPM.

First, we derive the non-dimensional formulation of the algebraic stress model and propose a machine learning framework for its implementation. Second, we employ symbolic regression techniques in conjunction with DNS datasets of periodic hill flows to train the Implicit-ASM. Third, we validate the model's performance using five test cases with distinctly different flow characteristics and compare its results with three alternative turbulence models. Our findings demonstrate that the Implicit-ASM performs effectively across all five test cases and exhibits notable advantages over the three comparative turbulence models.

For future research directions, we intend to develop a symbolic regression-based implicit algebraic stress model characterized by higher degrees of freedom.

\begin{acknowledgments}
This research is supported by the National Science and Technology Major Project of China (J2019-II-0005-0025).
\end{acknowledgments}

\section*{Data Availability Statement}

The data that support the findings of this study are available from the corresponding author upon reasonable request.

\bibliography{aipsamp}

\end{document}